\documentclass[11pt,draftcls,onecolumn,twoside]{IEEEtran} 
\usepackage[latin1]{inputenc}
\usepackage[mathcal]{euscript}
\usepackage{amssymb,amsmath,amsthm}
\usepackage{graphicx}
\usepackage[english]{babel}
\usepackage{textcomp}
\usepackage{pstricks,pst-plot,pst-text,pst-tree,pst-eps,pst-fill,pst-node,
pst-math}
\usepackage{algorithm}
\usepackage{algorithmic}

\newtheorem{remark}{Remark}
\newtheorem{prop}{Proposition}

\newtheorem{problem}{Problem}
\newtheorem{theo}{Theorem}
\newtheorem{lemma}{Lemma}
\newtheorem{conj}{Conjecture}

\usepackage{color}
\definecolor{MyDarkBlue}{rgb}{0.1,0,0.65}
\definecolor{MyDarkRed}{rgb}{.85,0,0.1}

\newcommand{\EE}[1]{{\mathbb{E}}\left[ #1 \right]}

\newcommand{\dsp}{\displaystyle}

\newcommand{\ki}{_{k,i}}
\newcommand{\ku}{_{k,1}}
\newcommand{\kd}{_{k,2}}

\newcommand{\lu}{_{l,1}}
\newcommand{\ld}{_{l,2}}
\newcommand{\Lu}{_{L,1}}

\newcommand{\Ld}{_{L,2}}

\title{Resource Allocation for Downlink Cellular OFDMA Systems:
Part I---Optimal Allocation}
\author{Nassar Ksairi$^{(1)}$\footnote{$^{(1)}$Sup\'elec, Plateau de Moulon
91192 Gif-sur-Yvette Cedex, France (nassar.ksairi@supelec.fr). Phone: +33 1 69
85 14 54, Fax: +33 1 69 85 14 69.}, Pascal Bianchi$^{(2)}$\footnote{$^{(2)}$CNRS
/ Telecom ParisTech (ENST), 46 rue Barrault 75634 Paris Cedex 13, France
(bianchi@telecom-paristech.fr,ciblat@telecom-paristech.fr,walid.hachem@enst.fr).
Phone: +33 1 45 81 83 60, Fax: +33 1 45 81 71 44.}, Philippe Ciblat$^{(2)}$,
Walid Hachem$^{(2)}$}

\begin{document}
\maketitle

\begin{abstract}
In this pair of papers (Part~I and Part~II in this issue), 
we investigate the issue of power control and
subcarrier assignment in a sectorized two-cell downlink OFDMA system impaired
by multicell interference. As recommended for WiMAX, we assume 
that the first part of the available bandwidth is likely to be
reused by different base stations (and is thus subject to multicell 
interference) and that the second part of the bandwidth is shared 
in an orthogonal way between the different base stations (and is 
thus protected from multicell interference).

Although the problem of multicell resource allocation is nonconvex 
in this scenario, we provide in Part~I the general form of the global solution.
In particular, the optimal resource allocation turns out to be ``binary'' in the
sense that, except for at most one pivot-user in each cell, any user receives
data either in the reused bandwidth or in the protected bandwidth, but not in
both. The determination of the optimal resource allocation essentially reduces
to the determination of the latter pivot-position.
\end{abstract}
\section{Introduction}

We consider the problem of resource allocation in the downlink of a sectorized
two-cell OFDMA system with incomplete Channel State Information (CSI) at the
Base Station (BS) side. In principle, performing resource allocation for
cellular OFDMA systems requires to solve the problem of power and subcarrier
allocation jointly in all the considered cells, taking into consideration the
interaction between users of different cells via the multicell interference.
Unfortunately, in most of the practical cases, this global optimization problem
is not convex and does not have, therefore, simple closed-form solution.
Practical alternative methods must thus to be proposed to perform the resource
allocation. Most of the works in the literature
on multicell resource allocation assumed perfect CSI on the transmitters side. 
In flat-fading scenarios with multi-user interference, a number of
interesting alternative methods have been proposed in the literature. One of
them is the \emph{geometric programming} (GP) approach proposed in \cite{gp} for
centralized power control scenarios. The author of this work showed that at high
SNR, the GP technique turns the nonconvex constrained optimization problem of
power control into a convex, thus tractable, optimization problem. Another
efficient resource allocation technique was proposed in \cite{boche1} for
decentralized power control scenarios. This technique is based on a min-max
formulation of the optimization problem, and is adapted to ad-hoc networks
contexts. Unfortunately, the two above mentioned techniques are mainly intended
for flat-fading scenarios, and are not directly suitable to general cellular
OFDMA contexts. To the best of our knowledge, only few works investigate OFDMA
multicell resource allocation. Authors of~\cite{scaling_laws} addressed the
optimization of the sum rate performance in a multicell network in order to
perform power control and user scheduling. In this context, the authors proposed
a decentralized algorithm that maximizes an upperbound on the network sum
rate. Interestingly, this upperbound is proved to be tight in the asymptotic
regime when the number of users per cell is allowed to grow to infinity.
However, the proposed  algorithm does not guaranty fairness among the different
users. 
In~\cite{bib_multicell3}, a centralized iterative allocation scheme
allowing to adjust the the number of cells reusing each subcarrier was
presented. The proposed algorithm does not suppose the so called ``reuse
partitioning'' scheme but nonetheless it promotes allocating subcarriers with
low reuse factors to users with bad channel conditions. It also provides an
interference limitation procedure in order to reduce the number of users whose
rate requirements is unsatisfied. Authors of~\cite{imp_reuse} considered the
problem of subcarrier assignment and power control that minimize the percentage
of unsatisfied users under rate and power constraints. For that sake, a
centralized algorithm based on reuse partitioning was proposed. In this
algorithm, the reuse factor of the far users next to the cell borders is adapted
according to the QoS  requirements and the problem parameters. Other dynamic
resource allocation schemes were proposed
in~\cite{algo_OFDMA1}-\cite{algo_OFDMA5}. The authors of~\cite{algo_OFDMA4}
and~\cite{algo_OFDMA5} have particularly discussed the issue of frequency  reuse
planning. It is worth mentioning here that neither of the above cited
works~\cite{bib_multicell3}-\cite{algo_OFDMA5}
provided analytical study of the performance of their respective proposed
schemes.
The issue of power control in distributed cooperative OFDMA networks was
addressed in~\cite{pischella}. However, the proposed solution assumes that
subcarrier allocation is performed independently from the power control.
The solution is thus suboptimal for the problem of resource allocation for
OFDMA networks, and a general solution for both power control and frequency
resource allocation remains to be provided. 

In contrast to previous works where perfect CSI was assumed, 
authors of~\cite{gau-hac-cib-1} assumed the knowledge of only the statistics of
users' channels and proposed an iterative algorithm for resource allocation in
the multicell context. In this algorithm a frequency (or subcarrier) reuse
factor equal to one was chosen, which means that each cell is supposed to use
all available subcarriers. This assumption relatively simplifies solving the
problem of multicell OFDMA resource allocation. A similar iterative multicell
allocation algorithm was proposed in~\cite{papandriopoulos} and its convergence
to the optimal solution of the multicell resource allocation problem was proved
based on the framework developed in~\cite{yates}.

In this paper, our aim is to characterize the resource allocation strategy 
(power control and subcarrier assignment scheme) allowing to satisfy all users'
rate requirements while spending the least power at the transmitters' side.
Similarly to~\cite{gau-hac-cib-1}, we investigate the case where the transmitter
CSI is limited to some channel statistics. However, contrary
to~\cite{gau-hac-cib-1} which assumes a frequency reuse factor equal to one, our
model assumes that a certain part of the available bandwidth is shared
orthogonally between the adjacent base stations (and is thus ``protected'' from
multicell interference) while the remaining part is reused by different base
stations (and is thus subject to multicell interference). 
Note that this so-called \emph{fractional frequency reuse} is recommended in a
number of standards \emph{e.g.} in~\cite{wimax2} for IEEE 802.16
(WiMax)~\cite{wimax1}. A similar reuse scheme is adopted in the recent
work~\cite{ofdma_game} which addresses the problem of power allocation in a
2-cell OFDMA system in order to maximize the system sum rate under a total power
constraint. The method proposed by the authors of~\cite{ofdma_game} to tackle
the latter problem is based on a game theory approach and it assumes that
subcarrier assignment is fixed in advance. 

As opposed to \cite{ofdma_game}, our work considers the problem of joint
optimization of power allocation and subcarrier assignment under the
aforementioned frequency reuse scheme. We also assume that each user is likely
to modulate in each of the two parts of the bandwidth (the protected and the non
protected parts). Thus, we stress the fact that \emph{i)} no user is forced to
modulate in a single frequency band, \emph{ii)} we do not assume \emph{a priori}
a geographical separation of users modulating in the two different bands. On the
opposite, we shall \emph{demonstrate} that such a geographical separation is
actually optimal w.r.t. our resource allocation problem. In this context, we
provide an algorithm that permits to compute the optimal resource allocation.

The paper is organized as follows. In Section~\ref{sec:sys_model} we present the
system model. In Section~\ref{sec:singlecell} we consider the problem of
resource allocation in a single cell assuming that the interference generated by
the other cells of the network is fixed. The problem consists in minimizing the
transmit power of the considered cell assuming a fixed level of interference
such that the rate requirements of users of this cell are satisfied and such
that the interference produced by the cell itself is less than a certain value.
Although resource allocation for users of the network requires in general
solving a multicell optimization problem, the single cell problem of
Section~\ref{sec:singlecell} turns out to be a useful tool to solve the more
complicated multicell problem. Theorem~\ref{the:single} gives the solution to
this single cell optimization problem. 
Except for at most one ``pivot'' user in the considered cell, 
any user receives data either in the interference bandwidth or in the protected
bandwidth, but not in both. In Section~\ref{sec:multicell} we introduce the
joint multicell resource allocation problem. This problem is equivalent to
jointly determining the resource allocation parameters of users belonging to
different interfering cells, such that all users' rate requirements are
satisfied and such that the total transmit power is minimized. 
Theorem~\ref{the:multi} characterizes the solution to this optimization problem
as function of a small number of unknown parameters. The solution turns out to
have in each cell the same binary form as the solution to the single cell
problem. Although this geographical separation is frequently used in practice,
no existing works prove the optimality of such a scheme to our knowledge. 
Subsection~\ref{sec:opt_algo} provides a method to calculate the optimal
resource allocation.
Finally, Section~\ref{sec:simus} is devoted to the numerical results.

\section{System Model}\label{sec:sys_model}
\subsection{OFDMA Signal Model}
\begin{figure}[h]
  \centering
  \includegraphics[width=10cm]{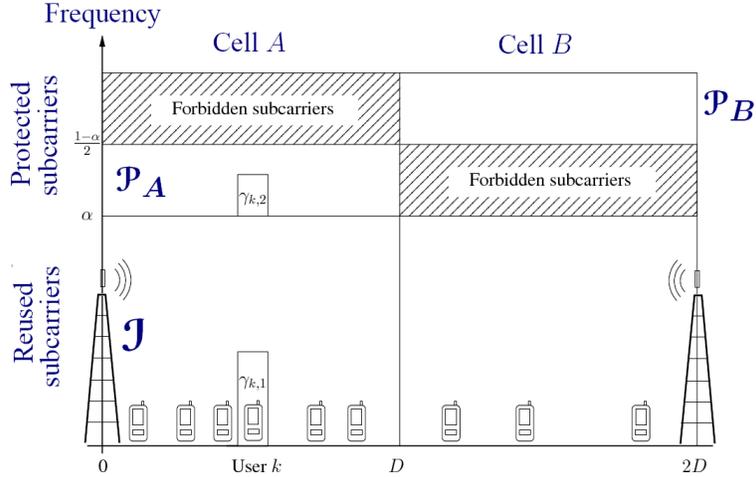}
  \caption{Two-Cell System model}
  \label{fig:model2}
\end{figure}

We consider a downlink OFDMA sectorized cellular network. In order to simplify
the presentation of our results, the network is supposed to be one-dimensional
(linear) as in a number of existing studies
\cite{gau-hac-cib-1,wyner,shamai,two_cell2,relay_1d}. The motivation behind our
choice of the one-dimensional network is that such a simple model can provide a
good understanding on the problem while still grasping the main aspects of a
real-world cellular system. It provides also some interesting guidelines that
help to implement practical cellular systems. Generalization to 2D-networks is
however possible (though much more involved) and is addressed in a separate
work~\cite{isita09}.
We consider the case of sectorized networks \emph{i.e.,} users
belonging to different sectors of the same cell are spatially
orthogonal~\cite{tse}.
In this case, it is reasonable to assume that a given user is only subject to
interference from the nearest interfering base station. 
Thus, we focus on two interfering sectors of two adjacent
cells, say Cell~$A$ and Cell~$B$, as illustrated by Figure~\ref{fig:model2}.
Denote by $D$ the radius of each cell which is assumed to be identical for all
cells without restriction. 
We denote by $K^A$ and $K^B$ the number of users in Cell $A$ and $B$
respectively. We denote by $K=K^A+K^B$ the total number of users in both cells.
Each base station provides information to all its users following a OFDMA
scheme. The total number of available subcarriers is denoted by $N$.
For a given user $k \in {1, 2, \ldots, K^A}$ in Cell $A$, we denote by 
${\cal N}_k$ the set of indices corresponding to the subcarriers modulated
by $k$. ${\cal N}_k$ is a subset of $\{0, 1, \ldots,N-1 \}$. By definition
of OFDMA, two distinct users $k, k'$ belonging to Cell $A$ are
such that ${\cal N}_k \cap {\cal N}_{k'} = \emptyset$. For each user 
$k \in \{1, \ldots, K^A \}$ of Cell $A$, the signal received by $k$ at the $n$th
subcarrier ($n \in {\cal N}_k$) and at the $m$th OFDM block is given by
\begin{equation}\label{eq:signal_model}
y_k(n,m) = H_k(n,m) s_k(n,m) + w_k(n,m), 
\end{equation} 
where $s_k(n,m)$ represents the data symbol transmitted by
Base Station~$A$. Process $w_k(n,m)$ is an additive noise
which encompasses the thermal noise and the possible multicell
interference.  Coefficient $H_k(n,m)$ 
is the frequency response of the channel at the subcarrier $n$
and the OFDM block $m$. Random variables $H_k(n,m)$ are
assumed to be Rayleigh distributed with variance
\begin{equation}\label{eq:variance}
\rho_k=\mathbb{E}[|H_{k}(m,n)|^{2}]\:.
\end{equation}
Note that the mean value $\rho_k$ does not depend on the subcarrier index. 
This is satisfied for instance in the case of decorrelated
channel taps in the time domain. For a given user $k$,
$H_k(n,m)$ are identically distributed w.r.t.
$n,m$, but are not supposed to be independent. Channel coefficients
are supposed to be perfectly known at the receiver side, and unknown at the base
station side. However, variances
$\rho_k$ are supposed to be known at the base station. This type of
incomplete CSI is particularly adapted to fast fading scenarios. In such a
context, sending feedback containing the instantaneous channel gain from users
to the base station will result in a significant overhead. 

As usual, we assume that $\rho_k$ vanishes with the distance between
Base Station~$A$ and user $k$, based on a given path loss model.
In the sequel, it is convenient to assume (without restriction)
that users $k = 1, 2, \ldots, K^A$ are numbered from the nearest to
the base station to the farthest. Therefore, for all users $k$ in Cell $A$,
\begin{equation}\label{eq:ordre_rho}
\rho_1>\rho_2>\ldots >\rho_{K^A}.
\end{equation}

\subsection{Frequency Reuse}
\label{sec:frequency_reuse}

The frequency reuse scheme is illustrated by Figure~\ref{fig:model2}. In
practical cellular OFDMA systems, it is usually assumed that certain
subcarriers $n \in \{0,\ldots N -1\}$ used by Base Station~$A$ are reused by the
adjacent Cell $B$. Denote by $\mathcal{I}$ this set of ``${\cal I}$nterfering''
subcarriers, $\mathcal{I} \subset \{0,\ldots, N-1\}$. If user $k$ modulates such
a subcarrier $n \in \mathcal{I}$, the additive noise $w_{k}(n,m)$ contains both
thermal noise of variance $\sigma^2$ and interference. Therefore, the variance
of $w_k(n,m)$ depends on $k$ and is crucially related to the position of user
$k$. We thus define for all $n \in \mathcal{I}$
\begin{equation*}
\mathbb{E}[|w_k(n,k)|^2]=\sigma_k^2\:.
\end{equation*}
Note that $\sigma_k^2$ is assumed to be a constant w.r.t. the subcarrier index
$n$. This assumption is valid in OFDMA multicell systems using
frequency hopping or random subcarrier assignment as in WiMax.
If users $k=1,2 \ldots K^A$ are numbered from the nearest to the base station to
the farthest, 
it is reasonable to assume that
\begin{equation}\label{eq:ordre_sigma}
\sigma_1^2<\sigma_2^2<\ldots <\sigma_{K^A}^2\:,
\end{equation}
meaning that the farthest users experience more multicell interference.
The \emph{reuse factor} $\alpha$ is defined as the ratio between the number of
reused subcarriers 
and the total number of available subcarriers:
\begin{equation*}
\alpha=\frac{\mbox{card}(\mathcal{I})}{N}
\end{equation*}
so that $\mathcal{I}$ contains $\alpha N$ subcarriers. The remaining
$(1-\alpha)N$ subcarriers 
are shared by the two cells, $A$ and $B$, in an orthogonal way. 
We assume that $\frac{1-\alpha}{2}N$ of these subcarriers 
are used by Base Station~$A$ only and are forbidden for $B$. 
Denote by $\mathcal{P}_A$ this set of ``$\mathcal{P}$rotected'' subcarriers. 
If user $k$ modulates such a subcarrier $n \in \mathcal{P}_A$, 
the additive noise $w_k(n,m)$ contains only thermal noise. 
In other words, subcarrier $n$ does not suffer from multicell interference. 
Then we simply write $\mathbb{E}[|w_k(n,m)|^2] = \sigma^2$, 
where $\sigma^2$ is the variance of the thermal noise only. 
Similarly, we denote by $\mathcal{P}_B$ the remaining $\frac{1-\alpha}{2}N$
subcarriers, 
such that each subcarrier $n \in \mathcal{P}_B$ is only used by Base
Station~$B$, and is not used by $A$. Finally, 
$\mathcal{I} \cup \mathcal{P}_A \cup \mathcal{P}_B= \{0, \ldots, N-1\}$.
Moreover, let $g_{k,1}$ (resp. $g_{k,2}$) be the channel 
Gain to Noise Ratio (GNR) in band $\mathcal{I}$ (resp. $\mathcal{P}_A$), 
namely $g\ku = \rho_k / \sigma_k^2$ (resp. $g\kd = {\rho_k} / {\sigma^2}$).

\subsection{Resource Allocation Parameters}
\label{sec:resource_parameters}

Of course, for a given user $k$ of Cell $A$, the noise variance $\sigma_k^2$
depends on the particular resource allocation used in the adjacent Cell $B$. 
We assume that $\sigma_k^2$ is known at Base Station~$A$, and that a given user
may use subcarriers in both the ``interference'' bandwidth $\mathcal{I}$ and the
``protected'' bandwidth $\mathcal{P}_A$. We denote by $\gamma\ku^AN$ (resp.
$\gamma\kd^AN$) the number of subcarriers modulated by user $k$ in the set
$\mathcal{I}$ (resp. $\mathcal{P}_A$). In other words,
\begin{equation*}
\gamma\ku^A = \mbox{card}(\mathcal{I} \cap \mathcal{N}_k)/N \qquad 
\gamma\kd^A = \mbox{card}(\mathcal{P}_A \cap \mathcal{N}_k)/N.
\end{equation*}
Note that by definition of $\gamma\ku^A$ and $\gamma\kd^A$,
$\sum_{k}\gamma\ku^A \leq \alpha$ and 
$\sum_k\gamma\kd^A \leq \frac{1-\alpha}{2}$, and that the superscript $A$ (or
$B$) is used to designate the cell in which user $k$ is located. We assume in
the sequel without restriction that the sharing factors
$\{\gamma\ku^A,\gamma\kd^A\}_{k}$ are continuous real-valued variables 
and can take on any value in the interval $[0,1]$. Furthermore, we assume that
a given user $k$ of Cell $A$ can modulate in both bands $\mathcal{I}$ and
$\mathcal{P}_A$ using distinct powers in each band. For any modulated subcarrier
$n \in \mathcal{N}_k$, we define 
$P\ku^A = E[|s_{k}(n,m)|^2]$ if $n\in \mathcal{I}$, 
$P\kd^A = E[|s_k(n,m)|^{2}]$ if $n\in \mathcal{P}_A$. 
Similarly, denote by $W\ki^A = \gamma\ki^A P\ki^A$ the average power transmitted
to user $k$ in $\mathcal{I}$ if $i = 1$ and in $\mathcal{P}_A$ if $i = 2$. 
``Setting a resource allocation for Cell $A$'' means setting a value 
for parameters $\{\gamma\ku^A, \gamma\kd^A, P\ku^A, P\kd^A\}_{k=1\ldots K^A}$,
or equivalently for parameters
$\{\gamma\ku^A, \gamma\kd^A,W\ku^A,W\kd^A\}_{k=1\ldots K^A}$.

\begin{remark}
As we stated above, the sharing factors $\gamma\ku^A$, $\gamma\kd^A$ are assumed
in our model to be \emph{real} numbers. This assumption does not necessarily
contradict the fact that each user can be assigned only \emph{integer} number of
subcarriers during the transmission of each OFDM symbol. Indeed, once the
real-valued $\{\gamma\ku^A,\gamma\kd^A\}_{k}$ are determined, the practical
subcarrier assignment can be done in several ways~\cite{gau-hac-cib-1}. One
possible way consists in allocating subcarriers to users according to some
frequency hopping pattern. In this case, the specific subset of subcarriers
assigned to each user varies from one OFDM symbol to another in such a way that
the average number of subcarriers modulated by each user $k$ in bands
$\mathcal{I}$ and $\mathcal{P}_A$ is equal to $\gamma\ku^A N$ and
$\gamma\kd^A N$ respectively. The latter frequency-hopping-based subcarrier
assignment scheme is assumed in this paper.
\end{remark}

\subsection{Multicell Interference Model}

We define now more clearly the way interference levels
$\sigma_1^2,\ldots,\sigma_{K^A}^2$ depend on the adjacent Base Station~$B$. In
OFDMA system models which assume frequency hopping like Flash-OFDM system
(\cite{tse} Chapter 4, page 179-180, \cite{flarion}), it is straightforward to
show that for a given user~$k$ of Cell~$A$, interference power $\sigma_k^2$ does
not depend on the particular resource allocation in Cell~$B$ but only on i) the
position of user~$k$ and ii) the average power $Q_1^B=\sum_{k=1}^{K^B}W\ku^B$
transmitted by Base Station~$B$ in the interference bandwidth~$\cal I$. 
More precisely,
\begin{equation}\label{eq:multicell_model}
\sigma_k^2=\mathbb{E}\left[|\tilde{H}_k(n,m)|^{2}\right]Q_1^B+\sigma^2
\end{equation}
where $\tilde{H}_k(n,m)$ represents the channel between Base Station~$B$ and
user~$k$ of Cell~$A$ at frequency~$n$ and OFDM block~$m$. In particular,
$\mathbb{E}\left[|\tilde{H}_k(n,m)|^{2}\right]$ only depends on the position of
user $k$ and on the path-loss exponent.

\section{Single Cell Resource Allocation}
\label{sec:singlecell}

Before tackling the problem of joint optimal resource allocation in 
the two considered cells, it is useful to consider first the simpler single cell
problem. The single cell formulation focuses on resource allocation in one cell,
and assumes that the resource allocation parameters of users 
in the other cell are fixed.

\subsection{Single Cell Optimization Problem}

Assume that each user $k$ has a rate requirement of $R_k$ nats/s/Hz. 
Our aim is to optimize the resource allocation for Cell $A$
which i) allows to satisfy all target rates $R_{k}$ of all users, 
and ii) minimizes the power used by Base Station~$A$ in order to achieve these
rates. 
Considering a fast fading context (i.e. channel coefficients $H_k(n,m)$ vary
w.r.t. $m$ all along the code word), 
we assume as usual that successful transmission at rate $R_k$ is possible
provided that $R_k < C_k$, 
where $C_k$ denotes the ergodic capacity associated with user $k$.  
Unfortunately, the exact expression of the ergodic capacity is difficult to
obtain
in our context due to the fact that the noise-plus-interference process 
$(w_k(n,m))_{n,m}$ is not a Gaussian process in general.
Nonetheless, if we endow the input symbols $s_k(n,m)$ with Gaussian
distribution, 
the mutual information between $s_k(n,m)$ and the received signal $y_k(n,m)$
in equation~\eqref{eq:signal_model} is minimum when the interference-plus-noise
$w_k(n,m)$ is Gaussian distributed.
Therefore, the approximation of the multicell interference as a Gaussian random
variable is widely used in the literature on OFDMA (see for instance
\cite{gau-hac-cib-1,coded_ofdma,noise_ofdma})
as it provides a lower bound on the mutual information.
For a given user~$k$ in Cell~$A$, the ergodic capacity in the whole band is
equal to the sum of the
ergodic capacities corresponding to both bands $\cal I$ and ${\cal P}_A$.
For instance, the part of the capacity corresponding to the protected band
${\cal P}_A$ is equal to
$\gamma\kd^A\EE{\log\left(1+P\kd^A\frac{|H_k(n,m)|^2}{\sigma^2}\right)}=
\gamma\kd^A\EE{\log\left(1+\frac{W\kd^A}{\gamma\kd^A}\frac{|H_k(n,m)|^2}{
\sigma^2}\right)}$, 
where factor $\gamma\kd^A$ traduces the fact that
the capacity increases with the number of subcarriers which are modulated by
user $k$. 
In the latter expression, the expectation is calculated with respect to random
variable $\frac{|H_k(m,n)|^2}{\sigma^2}$. 
Now, $\frac{H_k(m,n)|^2}{\sigma^2}$ has the same distribution as
$\frac{\rho_k}{\sigma^2}Z=g\kd Z$, where $Z$ is a standard exponentially
distributed random variable. 
Finally, the ergodic capacity in the whole bandwidth is equal to
\begin{equation}\label{eq:ergodic_capacity}
C_k(\gamma\ku^A,
\gamma\kd^A,W\ku^A,W\kd^A)=\gamma\ku^A\mathbb{E}\left[\log\left(1+g\ku\frac{
W\ku^A}{\gamma\ku^A}Z\right)\right]
+\gamma\kd^A\mathbb{E}\left[\log\left(1+g\kd\frac{W\kd^A}{\gamma\kd^A}
Z\right)\right]\:.
\end{equation}
The quantity $Q^A$ defined by
\begin{equation}\label{eq:average_power_K}
Q^A=\sum_{k=1}^{K^A}(W\ku^A+W\kd^A)
\end{equation}
denotes the average power spent by Base Station~$A$ during one OFDM block.
The optimal resource allocation problem for Cell~$A$ consists in characterizing
$\{\gamma\ku^A,\gamma\kd^A,W\ku^A,W\kd^A\}_{k=1\ldots K^A}$ 
allowing to satisfy all rate requirements of all
users ($R_k<C_k$) so that the power $Q^A$ to be spent is minimum.
Furthermore, as we are targeting a multicell interference scenario, 
it is also legitimate to limit the interference which is 
\emph{produced} by Base Station~$A$. Therefore, we introduce the following ``low
nuisance constraint'':
The power $Q_1^A=\sum_k W\ku^A$ which is transmitted by Base Station~$A$ in the
interference band $\cal I$
should not exceed a certain \emph{nuisance level} $\cal Q$, which is assumed to
be a predefined constant imposed
by the system's requirements. The introduction of this constraint will be later
revealed useful in Section~\ref{sec:multicell} when studying the solution to
the joint multicell resource allocation problem.
The single cell optimization problem can be formulated as follows.
\begin{problem}\label{prob:single}
Minimize $Q^A$ w.r.t. $\{\gamma\ku^A, \gamma\kd^A,W\ku^A,W\kd^A\}_{k=1\ldots
K^A}$ under the following constraints.
\begin{align*}
\mathbf{C1:}\: &\forall k,R_k\leq C_k & &\mathbf{C4:}\:
\gamma\ku^A\geq0,\gamma\kd^A\geq0\\
\mathbf{C2:}\: &\sum_{k=1}^{K^A}\gamma\ku^A = \alpha & &\mathbf{C5:}\:
W\ku^A\geq0,W\kd^A\geq0.\\
\mathbf{C3:}\: &\sum_{k=1}^{K^A}\gamma\kd^A = \frac{1-\alpha}{2} & & {\bf C6:}
\: \sum_{k=1}^{K^A} W\ku^A \leq {\cal Q}\:.
\end{align*}
\end{problem}
Here, $\bf C1$ is the rate constraint, $\bf C2$-$\bf C3$ are the bandwidth
constraints,
$\bf C4$-$\bf C5$ are the positivity constraints. Note that $\bf C6$ is the low
nuisance constraint imposed only on the power transmitted in the non protected
band $\cal I$. The particular case where the maximum admissible nuisance level
is set to ${\cal Q}=+\infty$ would correspond to a ``selfish'' resource
allocation: Base Station~$A$ may transmit as much power as needed in the
interference band $\cal I$ without caring about the nuisance which it produces
on the adjacent cell.
Note that in Problem~\ref{prob:single} no power constraint is imposed on the
total power $Q^A$ transmitted by the base station in the two bands.
Note also that the constraint set (the set of all feasible points) associated
with Problem~\ref{prob:single} is not empty as it contains at least the
following trivial solution. This trivial solution consists in assigning zero
power $W\ku^A=0$ on the subcarriers of the non protected band $\cal I$ (so that
constraint $\bf C6$ will be satisfied), and in performing resource allocation
only in the protected band $\mathcal{P}_A$.
The main reason for expressing the resource allocation problems in terms of
parameters $\gamma_{k,i}^A, W_{k,i}^A$ ($i=1,2$) instead of $\gamma_{k,i}^A,
P_{k,i}^A$ is that the ergodic capacity $C_k=C_k(\gamma_{k,1}^A,
W_{k,1}^A,\gamma_{k,2}^A, W_{k,2}^A)$ is a concave function of $\gamma_{k,i}^A,
W_{k,i}^A$.  As a consequence, the constraint set is convex and
Problem~\ref{prob:single} is a convex optimization problem in
$\{\gamma\ku^A,\gamma\kd^A,W\ku^A,W\kd^A\}_k$. Obviously, finding the optimal
parameter set $\{\gamma\ku^A, \gamma\kd^A,W\ku^A,W\kd^A\}_k$ is equivalent to
finding the optimal $\{\gamma\ku^A, \gamma\kd^A, P\ku^A, P\kd^A\}_k$ thanks to
the simple relation $W\ki^A=\gamma\ki^A P\ki^A$, $i=1,2$.

\subsection{Optimal Single Cell Resource Allocation}

In order to solve convex Problem~\ref{prob:single}, we use the Lagrange
Karush-Kuhn-Tucker (KKT) conditions. Define the following function on ${\mathbb
R}_+$ 
\begin{equation}\label{eq:functions_fF}
f(x) = \frac{\EE{\log(1+xZ)}}{\EE{\frac{Z}{1+xZ}}}-x\:.
\end{equation}
It can be shown that function $f(x)$ is increasing from 0 to
$\infty$ on ${\mathbb{R}}_+$. The following theorem provides the general form of
any global solution to Problem~\ref{prob:single}. Its proof is provided in
Appendix~\ref{app:main_theo}. 
\begin{theo}
\label{the:single} 
Any global solution $\{\gamma\ku^A, \gamma\kd^A,W\ku^A,W\kd^A\}_{k=1\ldots K^A}$
to Problem~\ref{prob:single} is as follows. There exists an integer
$L\in\{1,\dots, K^A\}$ and three nonnegative numbers $\beta_1,\beta_2$ and $\xi$
such that 
\begin{enumerate}
\item For each $k<L$,
\begin{equation}
\label{eq:allocinf}
	\begin{array}[h]{l|l}
	\dsp P\ku^A=g\ku^{-1}f^{-1}\left(\frac{g\ku}{1+\xi}\beta_1\right) &
	P\kd^A=0 \\
	\dsp \gamma\ku^A=\frac{R_k}{\EE{\log\left(1+ g\ku P\ku^A Z\right)}} &
\gamma\kd^A=0
	\end{array}
\end{equation}

\item For each $k>L$,
\begin{equation}
\label{eq:allocsup}
	\begin{array}[h]{l|l}
	 P\ku^A=0 & \dsp P\kd^A=g\kd^{-1}f^{-1}(g\kd \beta_2) \\
	\gamma\ku^A=0 & \dsp \gamma\kd^A=\frac{R_k}{\EE{\log\left(1+g\kd P\kd^A
Z\right)}}
	\end{array}
\end{equation}

\item For $k=L$
\begin{equation}
\label{eq:allocL}
	\begin{array}[h]{l|l}
	 \dsp P\ku^A=g\ku^{-1}f^{-1}\left(\frac{g\ku}{1+\xi}\beta_1\right) & 
	 \dsp P\kd^A=g\kd^{-1}f^{-1}(g\kd \beta_2) \\
	 \dsp \gamma\ku^A=\alpha-\sum_{l=1}^{k-1}\gamma\lu^A &
	 \dsp \gamma\kd^A=\frac{1-\alpha}{2}-\sum_{l=k+1}^{K^A}\gamma\ld^A,
	\end{array}
\end{equation}
\end{enumerate}
where $\beta_1$, $\beta_2$ and $\xi$ are the Lagrange multipliers associated
with constraints
$\bf C2$, $\bf C3$ and $\bf C6$ respectively. 
Determination of $L$,  $\beta_1$, $\beta_2$ and $\xi$ is provided by
Proposition~\ref{prop:param}.
\end{theo}

\noindent {\bf Comments on Theorem \ref{the:single}:}
\begin{enumerate}
\item[a)] Theorem \ref{the:single} states that the optimal resource allocation
scheme is ``binary'': Except for at most one user ($k = L$), any user receives
data either in the interference bandwidth $\mathcal{I}$ or in the protected
bandwidth $\mathcal{P}_A$, but not in both. Intuitively, it seems clear that
users who are the farthest from the base station should mainly receive data in
the protected bandwidth $\mathcal{P}_A$, as they are subject to an significant
multicell interference and hence need to be protected. Now, a closer look 
at our result shows that the farthest users should only receive in the protected
bandwidth $\mathcal{P}_A$. On the other hand, nearest users should only receive
in the interference bandwidth $\mathcal{I}$.

\item[b)] Nonzero resource allocation parameters $\gamma\ku^A, P\ku^A$ (for
$k\leq L$) and $\gamma\kd^A, P\kd^A$ (for $k\geq L$) are expressed as functions
of three parameters $\beta_1, \beta_2$,$\xi$. It can be easily seen from
Appendix~\ref{app:main_theo} that $\beta_1, \beta_2, \xi$ are the Lagrange
multipliers associated with constraints $\bf C2$, $\bf C3$ and $\bf C6$
respectively. It is quite intuitive that, when the admissible nuisance level is
large (take for instance $\cal Q=+\infty$), constraint $\bf C6$ holds with
strict inequality. Thus, $\xi=0$ from complementary slackness condition. In the
general case, the values of parameters $\beta_1, \beta_2, \xi$ can be obtained
from KKT conditions. The determination of $\beta_1, \beta_2, \xi$ and the
pivot-user $L$ is addressed in Subsection~\ref{sec:determination_L_beta}.

\item[c)] As expected, the optimal resource allocation depends on the resource
allocation in Cell~$B$ via parameters $\sigma_1^2,\ldots,\sigma_{K^A}^2$. Joint
optimization of the resource allocation in both cells, $A$ and $B$, is
investigated in Section \ref{sec:multicell}.
\end{enumerate}

\subsection{Determination of $L$, $\beta_1$, $\beta_2$  and $\xi$}
\label{sec:determination_L_beta}

{\bf Step 1: Determination of $\boldsymbol{L}$, $\boldsymbol{\beta_1}$,
$\boldsymbol{\beta_2}$ for a fixed value of $\boldsymbol{\xi}$.}

To simplify, first assume that the value of Lagrange multiplier $\xi$ is fixed.
We determine $L$, $\beta_1$, $\beta_2$ as functions of $\xi$.
Recall that user $L$ is defined as the only user who is likely to
modulate in both bands $\mathcal{I}$ and $\mathcal{P}_{A}$.
Parameters $\gamma_{L,1}^A, \gamma_{L,2}^A$ respectively provide the part of the
band $\mathcal{I}$ and $\mathcal{P}_{A}$ which is modulated by user $L$.
A first equation is obtained by writing that $C_L=R_L$ \emph{i.e.},
the rate constraint $\bf C1$ holds with equality. Recall that
$C_L$ is defined by~(\ref{eq:ergodic_capacity}) as 
$\gamma\Lu^A\EE{\log(1+g_{L,1}P\Lu^A Z)}+
\gamma\Ld^A\EE{\log(1+g_{L,2}P\Ld^A \xi)}$.
Define for each $x\geq 0$
\begin{equation}\label{eq:C_fun}
C(x) = \mathbb{E}[\log(1+f^{-1}(x)Z)]\:.
\end{equation}
Plugging the expression~(\ref{eq:allocL}) of parameters
$\gamma_{L,1}^A, P_{L,1}^A,\gamma_{L,2}^A, P_{L,2}^A$ into this expression,
equality $C_L/R_L=1$ becomes
\begin{equation}
\left[\alpha-\sum_{k<L}\frac{R_{k}}{C\left(\frac{g\ku}{1+\xi}{\beta}_{1}\right)}
\right]\frac{C\left(\frac{g\Lu}{1+\xi}{\beta}_{1}\right)}{R_L}+
\left[\frac{1-\alpha}{2}-\sum_{k>L}\frac{R_{k}}{C(g_{k,2}{\beta}_{2})}\right]
\frac{C(g_{L,2}{\beta}_{2})}{R_L}=1
\label{eq:userL2}
\end{equation}
In equation~(\ref{eq:userL2}), both terms enclosed inside the brackets coincide
with $\gamma\Lu^A$ and $\gamma\Ld^A$ respectively. As function $C(x)$ is
increasing from 0 to $\infty$ on ${\mathbb{R}}_+$,
constraints $\gamma\Lu^A\geq 0$ and $\gamma\Ld^A\geq 0$ hold only if
$\beta_1/(1+\xi)\geq a_{L-1}^A$ and $\beta_2\geq  b_L^A$
where for each $l$, $a_l^A$ and $b_l^A$ the unique positive numbers such that:
\begin{equation*}
\begin{array}{l@{\;\textrm{ and }\;}l}
\dsp\sum_{k=1}^{l}\frac{R_k}{C\left(g\ku a_l^A\right)}=\alpha 
&\dsp\sum_{k=l+1}^{K^A}\frac{R_k}{C(g\kd b_l^A)}=\frac{1-\alpha}{2},
\end{array}
\end{equation*}
with $a_0^A=b_{K^A}^A=0$ by convention. Note that $a_l^A$ is an increasing
sequence while $b_l^A$ is a decreasing sequence. Furthermore, in order
that~(\ref{eq:userL2}) holds, both (nonnegative) terms should be less than one.
Thus, 
$\dsp \alpha-\sum_{k\leq
L}\frac{R_k}{C\left(\frac{g\ku}{1+\xi}{\beta}_{1}\right)} \leq 0$
and $\dsp \frac{1-\alpha}{2}-\sum_{k\geq L}\frac{R_k}{C(g\kd{\beta}_2)} \leq 0$.
As a consequence, $\beta_1/(1+\xi)\leq a_L^A$ and $\beta_2\leq b_{L-1}^A$.
Finally,
\begin{equation}
\left(\frac{\beta_1}{1+\xi},\beta_2\right) \in [a_{L-1}^A, a_L^A] \times
[b_L^A, b_{L-1}^A ]\:.
\label{eq:domainebeta}
\end{equation}
Consider the case where $\gamma_{L,1}^A$, $\gamma_{L,2}^A$ are both nonzero,
and define the following function for each $x\geq 0$:
\begin{equation}\label{eq:functions_fF2}
F(x) = \EE{\frac{Z}{1+f^{-1}(x)Z}}\:.
\end{equation}
It can easily be seen from the KKT conditions derived in
Appendix~\ref{app:main_theo} that
\begin{equation}
\label{eq:gamma_non_nuls}
\frac{g\Lu}{1+\xi} F\left(\frac{g\Lu}{1+\xi}\beta_1\right)=g\Ld F(g\Ld\beta_2)
\:,
\end{equation}
Now using~(\ref{eq:domainebeta}) in the above equation along with the fact that
$F(.)$ is a decreasing function, one can easily see that $L$ can be defined as
\begin{equation}
\label{eq:L}
L = \min \left\{l=1\dots K^A \big/ 
\frac{g_{l,1}}{1+\xi}F\left(g_{l,1}a_l^A\right) \leq
g_{l,2}F\left(g_{l,2}b_l^A\right)\right\}.
\end{equation}
In practice, the search for $L$ can be achieved by dichotomy, 
computing $a_l^A$ and $b_l^A$ only for a limited number of values of $l$. 
Once $L$ is fixed, it is straightforward to show that the system formed by
equation
(\ref{eq:gamma_non_nuls}) and~(\ref{eq:userL2}) admits a unique solution
$(\beta_1,\beta_2)$. 
This is due to the fact that functions $C(.)$ and $F(.)$ are monotone.
Lagrange multiplier $\beta_1$, $\beta_2$ can thus be obtained using classical
root search tools.
As a remark, we note the existence of a rather pathological case, which
we do not address in details because of its limited importance.
To obtain equation~(\ref{eq:gamma_non_nuls}) we assumed that
$\gamma\Lu^A$ and $\gamma\Ld^A$ are strictly positive.
If this is not the case, say $\gamma\Lu^A=0$, 
it turns out that the system~(\ref{eq:userL2})-(\ref{eq:gamma_non_nuls})
has no solution. However, $L$ can still be obtained by~\eqref{eq:L} and 
$\beta_1,\beta_2$ can be easily obtained from~(\ref{eq:userL2}) 
which lead to $\beta_1=(1+\xi)a_L^A$, $\beta_2=b_{L-1}^A$. 
For the sake of simplicity, we will still refer to $(\beta_1, \beta_2)$
as the unique solution to system~(\ref{eq:userL2})-(\ref{eq:gamma_non_nuls}),
with slight language abuse, keeping in mind that we just put
$\beta_1=(1+\xi)a_L^A$, $\beta_2=b_{L-1}^A$
in the pathological case where such a solution does not exist. This convention
will be used throughout the paper without restriction.

\noindent {\bf Step 2: Determination of $\xi$.}

So far, we proved that for a fixed value of $\xi$, the optimal resource
allocation is unique and follows equations~(\ref{eq:allocinf}),
~(\ref{eq:allocsup}) and~(\ref{eq:allocL}), where $L=L(\xi)$ is given
by~(\ref{eq:L}) and $(\beta_1,\beta_2)=(\beta_1(\xi),\beta_2(\xi))$
is the unique solution to system~(\ref{eq:userL2})-(\ref{eq:gamma_non_nuls}).
The remaining task is now to determine $\xi$. Before addressing this point, it
is worth providing some insights on the impact of $\xi$ or equivalently, on the
role of the low nuisance constraint $\bf C6$ on the resource allocation.
Recall that $\xi$ is the Lagrange multiplier associated with constraint $\bf
C6$.
From an intuitive point of view, a large value of $\xi$ means in some sense that
constraint $\bf C6$ is severely restraining, whereas $\xi=0$ means that
constraint $\bf C6$ has no role and could have been deleted without modifying
the solution to Problem~\ref{prob:single}. It turns out that increasing $\xi$
has the effect of decreasing the total power $Q_1^A = \sum_k \gamma\ku^A P\ku^A$
which is transmitted in the interference band. This statement can be proved as
follows. First, we observe from equation~(\ref{eq:L})  that parameter $L=L(\xi)$
is a non increasing function of $\xi$. Second, it is straightforward to show
that for each $k$, $P\ku^A$ is a decreasing function of $\xi$.
Indeed, equation~(\ref{eq:allocinf}) implies that it is the composition of an 
increasing function $f^{-1}(x)$ and a decreasing function $\xi\mapsto
\beta_1(\xi)/(1+\xi)$  
(decreasingness of $\beta_1(\xi)/(1+\xi)$ is obtained after some algebra
from~(\ref{eq:userL2}) and~(\ref{eq:domainebeta})).
Third, $W\ku^A = P\ku^A R_k/\EE{\log(1+g\ku P\ku^A Z)}$ is an increasing
function of $P\ku^A$. It is thus a decreasing function of $\xi$ as a composition
of an increasing and a decreasing function $P\ku^A$. 
Therefore, the presence of an active constraint $\bf C6$ has a double impact on
the resource allocation: 
\emph{i)} it decreases the number $L$ of users who modulate in the interference
band $\cal I$, 
and \emph{ii)} it decreases the power $W\ku$ of each user in this band.

We now determine $\xi$. First we propose to compute the resource allocation
assuming $\xi=0$. If the corresponding value of $Q_1^A$ is such that $Q_1^A\leq
{\cal Q}$, then the procedure stops:
KKT conditions are met.
Otherwise, this means that constraint $\bf C6$ should be active: $\xi >0$. From
complementary slackness condition, 
$\bf C6$ should be met with equality : one should determine $\xi$ such that
$Q_1^A=\sum_k \gamma\ku^AP\ku^A$ coincides with $\cal Q$:
\begin{equation}
  \label{eq:C6egalite}
  \sum_{k\leq L} \gamma\ku^AP\ku^A = {\cal Q},
\end{equation}
where $\gamma\ku^A, P\ku^A$ are defined by~(\ref{eq:allocinf}) and where
$L=L(\xi)$, $\beta_1=\beta_1(\xi)$, $\beta_2=\beta_2(\xi)$
have been defined previously. As mentioned above, $Q_1^A$ is a decreasing
function of $\xi$ so that the solution $\xi$ to equation $Q_1^A = {\cal Q}$ is
unique. 

Finally, we conclude that the following proposition holds true.
\begin{prop}
  \label{prop:param}
The global solution to the single cell Problem~\ref{prob:single} is unique and
is given by equations~(\ref{eq:allocinf})-(\ref{eq:allocsup})-(\ref{eq:allocL}),
where parameters $L$, $\beta_1$, $\beta_2$ and $\xi$ are unique and determined
as follows. 
\begin{enumerate}
\item Assuming $\xi=0$, evaluate $L$ by (\ref{eq:L}) and $(\beta_1, \beta_2)$
as the unique solution to the system of
equation~(\ref{eq:userL2})-(\ref{eq:gamma_non_nuls}) satisfying
$\left(\frac{\beta_1}{1+\xi},\beta_2\right) \in [a_{L-1}^A, a_L^A] \times
[b_L^A, b_{L-1}^A ]$. Then evaluate $Q_1^A= \sum_k \gamma\ku^A P\ku^A$.
\item Stop if $Q_1^A \leq {\cal Q}$ (constraint $\bf C6$ is met) otherwise
continue.
\item Evaluate ($L$, $\beta_1$, $\beta_2$, $\xi$) as the unique solution to the
system of equations
(\ref{eq:userL2})-(\ref{eq:gamma_non_nuls})-(\ref{eq:L})-(\ref{eq:C6egalite}).
\end{enumerate}
\end{prop}
The above proposition proves that the global solution to the single cell
allocation problem is unique and provides a procedure to compute it.
Algorithm ~\ref{algo:single_cell} gives a more detailed description of the
latter procedure and proposes a method to solve the system of equation
(\ref{eq:userL2})-(\ref{eq:gamma_non_nuls})-(\ref{eq:L})-(\ref{eq:C6egalite}).
\begin{algorithm}
\caption{Determination of $L$, $\beta_1$, $\beta_2$, $\xi$}
\label{algo:single_cell}
\begin{algorithmic}
\STATE $\xi \leftarrow 0$
\REPEAT
\STATE $L \leftarrow \min \left\{l \big/ 
\frac{g_{l,1}}{1+\xi}F\left(g_{l,1}a_l^A\right) \leq
g_{l,2}F\left(g_{l,2}b_l^A\right)\right\}$
\STATE $(\beta_1,\beta_2)\leftarrow$ unique solution in 
$[a_{L-1}^A, a_L^A] \times [b_L^A, b_{L-1}^A ]$ to
\eqref{eq:userL2}-\eqref{eq:gamma_non_nuls} 
\STATE $Q_1^A \leftarrow \sum_{k=1}^{L-1}
\frac{R_k}{C\left(\frac{g\ku}{1+\xi}\beta_1\right)}
g\ku^{-1}f^{-1}\left(\frac{g\ku}{1+\xi}\beta_1\right)+
\left(\alpha-\sum_{k=1}^{L-1}\frac{R_k}{C\left(\frac{g\ku}{1+\xi}\beta_1\right)}
\right)g_{L,1}^{-1}f^{-1}\left(\frac{g_{L,1}}{1+\xi}\beta_1\right)$
\IF{$Q_1^A > \mathcal{Q}$}
\STATE Increment $\xi$
\ENDIF
\UNTIL{$Q_1^A \leq \mathcal{Q}$}
\RETURN $L$, $\beta_1$, $\beta_2$, $\xi$
\end{algorithmic}
\end{algorithm}

One still needs to define the way $\xi$ should be incremented at the end of
each iteration of Algorithm~\ref{algo:single_cell}. In practice, updating the
value of $\xi$ can be done by resorting to Newton-like methods which are widely
used to solve nonlinear equations.

\section{Joint Multicell Resource Allocation}
\label{sec:multicell}

\subsection{Optimization Problem}

Our aim now is to jointly optimize the resource allocation for the two cells
which i) allows to satisfy all target rates $R_k$ of all users, 
and ii) minimizes the power used by the two base stations in order to achieve
these rates. 
The ergodic capacity associated with user~$k$ in Cell~$A$ is given by
equation~\eqref{eq:ergodic_capacity},
where coefficient $g\ku$ in that equation coincides with
$$
g\ku(Q_{1}^B)=\frac{\rho_k}{\mathbb{E}\left[|\tilde{H}_k(n,m)|^{2}\right]
Q_1^B+\sigma^2},
$$
where $\tilde{H}_k(n,m)$ represents the channel between Base Station~$B$ and
user~$k$ of Cell~$A$ at frequency~$n$ and OFDM block~$m$. Coefficient
$g\ku(Q_{1}^B)$ represents user $k$ channel gain to interference-plus-noise
ratio in the interference band $\cal I$. Here, $g\ku(Q_{1}^B)$ not only depends
on the position of user $k$ in Cell $A$, but also on the power
$Q_1^B=\sum_{k=1}^{K^B}W\ku^B$ transmitted by the adjacent Base Station~$B$ in
band~$\cal I$. We now solve the following multicell resource allocation
problem. 
\begin{problem}\label{prob:multi}
Minimize the total power spent by both base stations
$Q=\dsp \sum_{c=A,B}\sum_{k=1}^{K^c}(W\ku^c+W\kd^c)$ with respect to
$\{\gamma\ku^c, \gamma\kd^c,W\ku^c,W\kd^c\}_{\substack{c=A,B\\ k=1\ldots K^c}}$ 
under the following constraints.
\begin{align*}
\mathbf{C1:}\: &\forall k,R_k\leq C_k & &\mathbf{C4:}\:
\gamma\ku^c\geq0,\gamma\kd^c\geq0\\
\mathbf{C2:}\: &\sum_{k=1}^{K^c}\gamma\ku^c=\alpha & &\mathbf{C5:}\:
W\ku^c\geq0,W\kd^c\geq0.\\
\mathbf{C3:}\: &\sum_{k=1}^{K^c}\gamma\kd^c=\frac{1-\alpha}{2}
\end{align*}
\end{problem}
It can be easily seen that the above optimization problem is feasible as soon as
$\alpha < 1$. Indeed, a naive but nevertheless feasible point can be easily
constructed by forcing each user to modulate in the protected band only (force
$\gamma\ku^c=0$ for each user). Cells thus become orthogonal, and all users rate
requirements $R_k$ can be satisfied provided that enough power is transmitted in
the protected band.
Unfortunately, the ergodic capacity $C_k$ of user~$k$ is not a convex function
with respect to the optimization variables. This is due to the fact that the
gain-to-noise ratio $g\ku(Q_1^B)$ is a function of the resource allocation
parameters of users belonging to the interfering cell. Therefore, Problem
\ref{prob:multi} is nonconvex, and cannot be solved by classical convex
optimization methods. Nonetheless, we manage to characterize its solution. 
In fact, we prove that the solution has the same simple binary form of the
single cell optimal solution.

\subsection{Optimal Resource Allocation}

For each cell $c\in\{A,B\}$, denote by $\overline{c}$ the adjacent cell
($\overline{A}=B$ and $\overline{B}=A$).
The following result is proved in Appendix~\ref{sec:multi_proof}.
\begin{theo}
\label{the:multi}
~\\{\bf(A)} Any global solution to Problem~\ref{prob:multi} has the following
form.
For each Cell $c$, there exists an integer $L^c\in\{1,\dots, K^c\}$, and there
exist four positive numbers $\beta_1^c$, $\beta_2^c$, $\xi^c$, $Q_1^{\bar c}$ 
such that 
\begin{enumerate}
\item For each $k<L^c$,
\begin{equation}
\label{eq:multiinf}
	\begin{array}[h]{l|l}
	 \dsp P\ku^c={g\ku(Q_1^{\bar c})}^{-1}f^{-1}\left(\frac{g\ku(Q_1^{\bar
c})}{1+\xi^c}\beta_1^c\right) & P\kd^c=0 \\
	\dsp \gamma\ku^c=\frac{R_k}{\EE{\log\left(1+{g\ku(Q_1^{\bar c})} P\ku^c
Z\right)}} & \gamma\kd^c=0
	\end{array}
\end{equation}

\item For each $k>L^c$,
\begin{equation}
\label{eq:multisup}
	\begin{array}[h]{l|l}
	 P\ku^c=0 & \dsp P\kd^c=g\kd^{-1}f^{-1}(g\kd \beta_2^c) \\
	\gamma\ku^c=0 & \dsp \gamma\kd^c=\frac{R_k}{\EE{\log\left(1+g\kd P\kd^c
Z\right)}}
	\end{array}
\end{equation}

\item For $k=L^c$
\begin{equation}
\label{eq:multiL}
	\begin{array}[h]{l|l}
	 \dsp P\ku^c={g\ku(Q_1^{\bar c})}^{-1}f^{-1}\left(\frac{g\ku(Q_1^{\bar
c})}{1+\xi^c} \beta_1\right) & 
	 \dsp P\kd^c=g\kd^{-1}f^{-1}(g\kd \beta_2^c) \\
	 \dsp \gamma\ku^c=\alpha-\sum_{l=1}^{k-1}\gamma\lu^c &
	 \dsp \gamma\kd^c=\frac{1-\alpha}{2}-\sum_{l=k+1}^{K^c}\gamma\ld^c.
	\end{array}
\end{equation} 
\end{enumerate}
{\bf(B)} For each $c=A,B$, the system ${\cal S}^c(Q_1^A,Q_1^B)$ formed by the
following four equations is satisfied.
\begin{gather}
  L^c = \min \left\{l=1\dots K^c \big/ \frac{g\lu(Q_1^{\bar c})}{1+\xi^c}
  F\left(\frac{g\lu(Q_1^{\bar c})}{1+\xi^c}a_{l}\right) \leq
g_{l,2}F\left(g_{l,2}b_l\right)\right\} 
\label{eq:Lmulti}\\
  \frac{g_{L^c,1}(Q_1^{\bar c})}{1+\xi^c} 
  F\left(\frac{g_{L^c,1}(Q_1^{\bar c})}{1+\xi^c} \beta_1^c\right)=g_{L^c,2}
F(g_{L^c,2}\beta_2^c) 
\label{eq:FFmulti}\\
  \gamma_{L^c,1}^c C\left(\frac{g_{L^c,1}(Q_1^{\bar
c})}{1+\xi^c}\beta_1^c\right)+
  \gamma_{L^c,2}^c C(g_{L,2}{\beta}_{2}^c)=R_{L^c}
\label{eq:RLCLmulti}\\
	\sum_k^{L^c}\gamma\ku^c P\ku^c= Q_1^c\:,
\label{eq:Qmulti}
\end{gather}
where the values of $\gamma\ku^c$ and $P\ku^c$ in~(\ref{eq:Qmulti}) are the
functions of $(\beta_1^c,\beta_2^c,\xi^c)$ defined by
equation~(\ref{eq:multiinf}).\\
{\bf(C)} Furthermore, for each $c=A,B$ and for any arbitrary values
$\tilde{Q}_1^A$ and $\tilde{Q}_1^B$, the system of equations 
${\cal S}^c(\tilde{Q}_1^A,\tilde{Q}_1^B)$ admits at most one solution
$(L^c,\beta_1^c,\beta_2^c,\xi^c)$.
\end{theo}
\medskip

\noindent {\bf Comments on Theorem~\ref{the:multi}:}

\begin{enumerate}
\item[a)] The joint multicell resource allocation problem required initially 
the determination of $4K$ parameters (where $K$ is the total number of users). 
Theorem~\ref{the:multi} allows to reduce the search to only two parameters,
namely $Q_1^A$ and $Q_1^B$.  Once the value of these parameters is fixed, the
resource allocation parameters for each user can be obtained from the above
results. As a consequence, the only remaining task is to determine the value of
$(Q_1^A,Q_1^B)$. This task is addressed in Subsection~\ref{sec:opt_algo}.

\item[b)] We observe that the system ${\cal S}^c(Q_1^A,Q_1^B)$ is very similar
to the system obtained in the single cell case at equations~\eqref{eq:userL2},
\eqref{eq:gamma_non_nuls}, \eqref{eq:L} and \eqref{eq:C6egalite}. In fact, as
stated by the proof later, the optimal resource allocation in the multicell case
can be interpreted as the solution to a certain single-cell problem.

\item[c)] As a consequence of the above remark, Theorem~\ref{the:multi} states
that the optimal multicell resource allocation scheme has the same ``binary''
form as in the single cell case. Even if optimal resource allocation is achieved
\emph{jointly} for both interfering cells, there still exists a pivot-user $L^c$
in each Cell~$c$ which separates the users modulating respectively in bands
$\cal I$ and ${\cal P}_c$.
 
\item[d)] It is worth noticing that this binary resource allocation strategy is
already proposed in a number of recent standards. One of the contributions
introduced by Theorem~\ref{the:multi} is the proof that such a strategy in not
only simple and intuitive, but is also optimal.
\end{enumerate}

\subsection{Optimal Distributed Algorithm}
\label{sec:opt_algo}

Once the relevant values of $Q_1^A$ and $Q_1^B$ have been determined, each base
station can easily compute the optimal resource allocation based on
Theorem~\ref{the:multi}. As a consequence, the only remaining task is to
determine the value of $(Q_1^A,Q_1^B)$. To that end, we propose to perform an
exhaustive search on~$(Q_1^A,Q_1^B)$.

\noindent {\sl i)} For each point $(\tilde Q_1^A,\tilde Q_1^B)$ on a certain
2D-grid (whose determination will be discussed later on), each base station
$c=A,B$ solves the system ${\cal S}^c(\tilde Q_1^A,\tilde Q_1^B)$ 
introduced by Theorem~\ref{the:multi}.
Solving ${\cal S}^c(\tilde Q_1^A,\tilde Q_1^B)$ for arbitrary values 
$(\tilde Q_1^A, \tilde Q_1^B)$ can be easily achieved by base station $c$ thanks
to a simple {\sl single-cell} procedure. 
Focus for instance on Cell~$A$.
\begin{itemize}
\item Base station $A$ solves the single cell 
resource allocation Problem~\ref{prob:single}
assuming that the interference level coincides with 
$\tilde Q_1^B$, and that the nuisance constraint ${\cal Q}$ is set to 
${\cal Q}=\tilde Q_1^A$. 
The (unique) solution is provided by Theorem~\ref{the:single} and
Proposition~\ref{prop:param}.

\item If the resulting power $\sum_k \gamma\ku^AP\ku^A$ transmitted 
in the interference band ${\cal P}_A$
is equal to the nuisance constraint $\tilde Q_1^A$ ({\sl i.e.} 
constraint $\bf C6$ holds with equality),
then the resulting value of $(L^A,\beta_1^A,\beta_2^A,\xi^A)$ 
coincides with the unique solution
to system ${\cal S}^A(\tilde Q_1^A,\tilde Q_1^B)$.
This claim is the immediate consequence of Proposition~\ref{prop:param}.

\item If the power $\sum_k \gamma\ku^AP\ku^A$ is less than $\tilde Q_1^A$
({\sl i.e.} constraint $\bf C6$ holds with strict inequality), 
then $(L^A,\beta_1^A,\beta_2^A,\xi^A)$ is clearly not a solution to system 
${\cal S}^A(\tilde Q_1^A,\tilde Q_1^B)$,
as equality~\eqref{eq:Qmulti} does not hold. In this case, it can easily
be seen that ${\cal S}^A(\tilde Q_1^A,\tilde Q_1^B)$ has no solution.
The point $(\tilde Q_1^A,\tilde Q_1^B)$ cannot correspond to a global solution
as stated by Theorem~\ref{the:multi} and is thus eliminated.
\end{itemize}

\noindent {\sl ii)} Base station A evaluates the power
$$
Q_T^A(\tilde Q_1^A,\tilde Q_1^B) = \sum_k \gamma\ku^A P\ku^A+\gamma\kd^A P\kd^A
$$
that would be transmitted if the interference level and the nuisance constraint
were respectively equal to $\tilde Q_1^B$ and $\tilde Q_1^A$. This value is then
communicated to Base Station~B. Base station B proceed in a similar way.

\noindent {\sl iii)} The final value of $(Q_1^A,Q_1^B)$ is defined as the
argument of the minimum power
transmitted by the network:
$$
(Q_1^A,Q_1^B) = \arg\min_{(\tilde Q_1^A,\tilde Q_1^B)} Q_T^A(\tilde Q_1^A,\tilde
                 Q_1^B)+ Q_T^B(\tilde Q_1^A,\tilde Q_1^B)\:.
$$
\medskip

Note that the optimal resource allocation algorithm as described above does not
require the intervention of a central controlling unit supposed to have access
to the two base stations and to users' information (position and data rate). We
only assume that both base stations can communicate via a special link dedicated
to this task. The algorithm is thus distributed. This special link will be only
used to exchange a limited number of messages. Indeed, the only values that need
to be exchanged between the two base stations are $Q_T^A(\tilde Q_1^A,\tilde
Q_1^B)$ and $Q_T^B(\tilde Q_1^A,\tilde Q_1^B)$ corresponding to the couples
$(\tilde Q_1^A,\tilde Q_1^B)$ for which the two systems of equations ${\cal
S}^A(\tilde Q_1^A,\tilde Q_1^B)$ and ${\cal S}^B(\tilde Q_1^A,\tilde Q_1^B)$
have a solution. Algorithm~\ref{algo:multi_cell} given below summarizes the
steps involved in the optimal resource allocation.
\begin{algorithm}
\caption{Optimal distributed allocation algorithm}
\label{algo:multi_cell}
\begin{algorithmic}
\STATE {\bf 1. Single cell processing}
\STATE
\begin{tabular}{l|l}
{\bf Cell $\boldsymbol{A}$} &{\bf Cell $\boldsymbol{B}$} \\
\hline
{\bf for} each $(\tilde{Q}_1^A,\tilde{Q}_1^B)$ {\bf do}&
{\bf for} each $(\tilde{Q}_1^A,\tilde{Q}_1^B)$ {\bf do}\\
\:\:\:\:$\{\gamma\ki^A,P\ki^A\}_{i,k}\leftarrow$ Solve ${\cal S}^A(\tilde
Q_1^A,\tilde Q_1^B)$ using &
\:\:\:\:$\{\gamma\ki^B,P\ki^B\}_{i,k}\leftarrow$ Solve
${\cal S}^B(\tilde Q_1^A,\tilde Q_1^B)$ using\\
\:\:\:\:Algorithm~\ref{algo:single_cell} with $\mathcal{Q}=\tilde{Q}_1^A$ &
\:\:\:\:Algorithm~\ref{algo:single_cell} with $\mathcal{Q}=\tilde{Q}_1^B$\\
{\bf if} $\sum_{k=1}^{K^A}\gamma\ku^A P\ku^A=\tilde{Q}_1^A$ {\bf then} &
{\bf if} $\sum_{k=1}^{K^B}\gamma\ku^B P\ku^B=\tilde{Q}_1^B$ {\bf then}\\
\:\:\:\:$Q_T^A(\tilde{Q}_1^A,\tilde{Q}_1^B)\leftarrow$
$\sum_{i=1,2}\sum_{k=1}^{K^A}\gamma\ki^A P\ki^A$ &
\:\:\:\:$Q_T^B(\tilde{Q}_1^A,\tilde{Q}_1^B)\leftarrow$
$\sum_{i=1,2}\sum_{k=1}^{K^B}\gamma\ki^B P\ki^B$\\
{\bf end if} &{\bf end if}\\
{\bf end for} &{\bf end for}
\end{tabular}
\STATE {\bf 2. Cooperation between BS $\boldsymbol{A}$ and $\boldsymbol{B}$}
\STATE $(Q_1^A,Q_1^B) \leftarrow \arg\min_{(\tilde Q_1^A,\tilde Q_1^B)}
Q_T^A(\tilde Q_1^A,\tilde Q_1^B)+ Q_T^B(\tilde Q_1^A,\tilde Q_1^B)$
\STATE {\bf 3. Resource allcoation in each cell}
\STATE
\begin{tabular}{l|l}
{\bf Cell $\boldsymbol{A}$} &{\bf Cell $\boldsymbol{B}$} \\
\hline
$\{\gamma\ki^A,P\ki^A\}_{i,k}\leftarrow$ Solve 
${\cal S}^A(Q_1^A,Q_1^B)$ using\:\:\:\:\:\:&
$\{\gamma\ki^B,P\ki^B\}_{i,k}\leftarrow$ Solve
${\cal S}^B(Q_1^A, Q_1^B)$ using\:\:\:\:\:\:\\
Algorithm~\ref{algo:single_cell} with $\mathcal{Q}=Q_1^A$ &
Algorithm~\ref{algo:single_cell} with $\mathcal{Q}=Q_1^B$
\end{tabular}
\end{algorithmic}
\end{algorithm}

\noindent {\bf Determination of the search domain in $\boldsymbol{(Q_1^A,
Q_1^B)}$.}

In order to limit the complexity of the proposed approach, the search for
$(Q_1^A,Q_1^B)$ should be restricted to a certain compact domain, say
$$
Q_1^c \in [0, {\cal Q}_{max}]
$$
for each $c$. For instance, a possible value for ${\cal Q}_{max}$ can be defined as
the total power that would be spent by the two base stations if one would use
the naive and suboptimal 
resource allocation which consists in only transmitting in the protected bands 
${\cal P}_A$ and ${\cal P}_B$.
Clearly, the latter value of ${\cal Q}_{max}$ is a constant w.r.t. $Q_1^A$ and $Q_1^B$
and can be computed beforehand.
A second way to restrict the search domain is to make use of a simple suboptimal
multicell resource allocation algorithm prior to the use of our algorithm
(see for instance the suboptimal algorithm defined in Part~II of this work).
In this case, it is possible to restrict the search for $(Q_1^A,Q_1^B)$ to a 
well-chosen neighborhood of the couple $(Q_1^A,Q_1^B)_{\textrm{subopt}}$
provided by the suboptimal solution. 

\noindent {\bf Complexity Analysis.} 

In order to get an idea about the cost of applying the optimal allocation, we
provide in the following a computational complexity analysis of this algorithm
as function of the number of users $K$ in the system. For that sake, recall that
the system of equations 
${\cal S}^c(\tilde{Q}_1^A,\tilde{Q}_1^B)$ must be solved for each possible value
of $(\tilde{Q}_1^A,\tilde{Q}_1^B)$ inside a 2D-grid contained in a compact
interval. 

For each point $(\tilde{Q}_1^A,\tilde{Q}_1^B)$ of the 2D-grid, solving 
${\cal S}^c(\tilde{Q}_1^A,\tilde{Q}_1^B)$ can be done by a procedure
similar to Algorithm~\ref{algo:single_cell}.
Recall that during each iteration of the latter algorithm, the value of $L^c$
should be determined by solving $L^c=\min\big\{l=1\dots K^c \big/$
$\frac{g_{l,1}}{1+\xi^c} F\left(g_{l,1}a_l^c\right) \leq
g_{l,2}F\left(g_{l,2}b_l^c\right)\big\}$. This requires that a certain subset
of parameters $\{a_l^c, b_l^c\}$ should be computed first.
It can be shown that the number of operations needed to compute $a_l^c, b_l^c$
is of order $O(K^c)$. Furthermore, we argued in
Subection~\ref{sec:determination_L_beta} that the determination of $L^c$ can be
done by dichotomy, computing $a_l^c$ and $b_l^c$ only for a limited number,
namely $\log_2 K^c$, of values of $l$. The overall complexity of finding $L^c$
for a fixed $\xi^c$ is therefore of the order of $O(K^c \log_2 K^c)$. 

Once $L^c$ is determined, the following step of Algorithm~\ref{algo:single_cell}
consists in solving the system of
equation~\eqref{eq:userL2}-\eqref{eq:gamma_non_nuls} in variables
$\beta_1^c,\beta_2^c$. This system of non linear equations can be solved using
Newton-like iterative methods. One can verify by referring to~\cite{non_linear}
that the latter system requires a computational complexity proportional to
$O(K^c)$. The computational complexity associated with each iteration of
Algorithm~\ref{algo:single_cell} is therefore dominated by the cost of computing
$L^c$, which is of order $O(K^c \log_2 K^c)$. Denote by $N_i$ the number of
iterations of Algorithm~\ref{algo:single_cell} needed till convergence. We
conclude that the overall computational complexity of solving
$\mathcal{S}^c(\tilde{Q}_1^A,\tilde{Q}_1^B)$ is of the order of 
$O(N_i K^c \log_2 K^c)$.

Denote by $M$ the number of couples $(\tilde{Q}_1^A,\tilde{Q}_1^B)$ in
the 2D-grid. The overall computational complexity of the algorithm can be
obtained by multiplying the cost of solving 
${\cal S}^c(\tilde{Q}_1^A,\tilde{Q}_1^B)$ for each point of the 2D-grid by $M$
the number of points in the grid. The latter overall cost is therefore of the
order of $O(M N_i K^A \log_2 K^A) + O(M N_i K^B
\log_2 K^B)$, which is itself of order $O(M N_i K \log_2 K)$ in the particular
case $K^A\sim K^B\sim K/2$. 

Note from the above discussion that the determination of the pivot-user $L^c$ 
in each cell for each value of $(Q_1^A,Q_1^B)$ is one of the costliest
operations in solving ${\cal S}^c(Q_1^A,Q_1^B)$ and that it dominates the
overall complexity.
This is why we propose in Part~II of this work a simplified resource allocation
algorithm which uses a predefined value for the pivot distance.
The simplified algorithm turns out to have a computational complexity
of order $O(K)$, as opposed to the computational complexity of the optimal
algorithm which is of the order of $O(M N_i K \log_2 K)$.

\section{Simulations}
\label{sec:simus}

\begin{figure}[h]
   \begin{minipage}[b]{0.49\linewidth}
      \centering \includegraphics[scale=0.25]{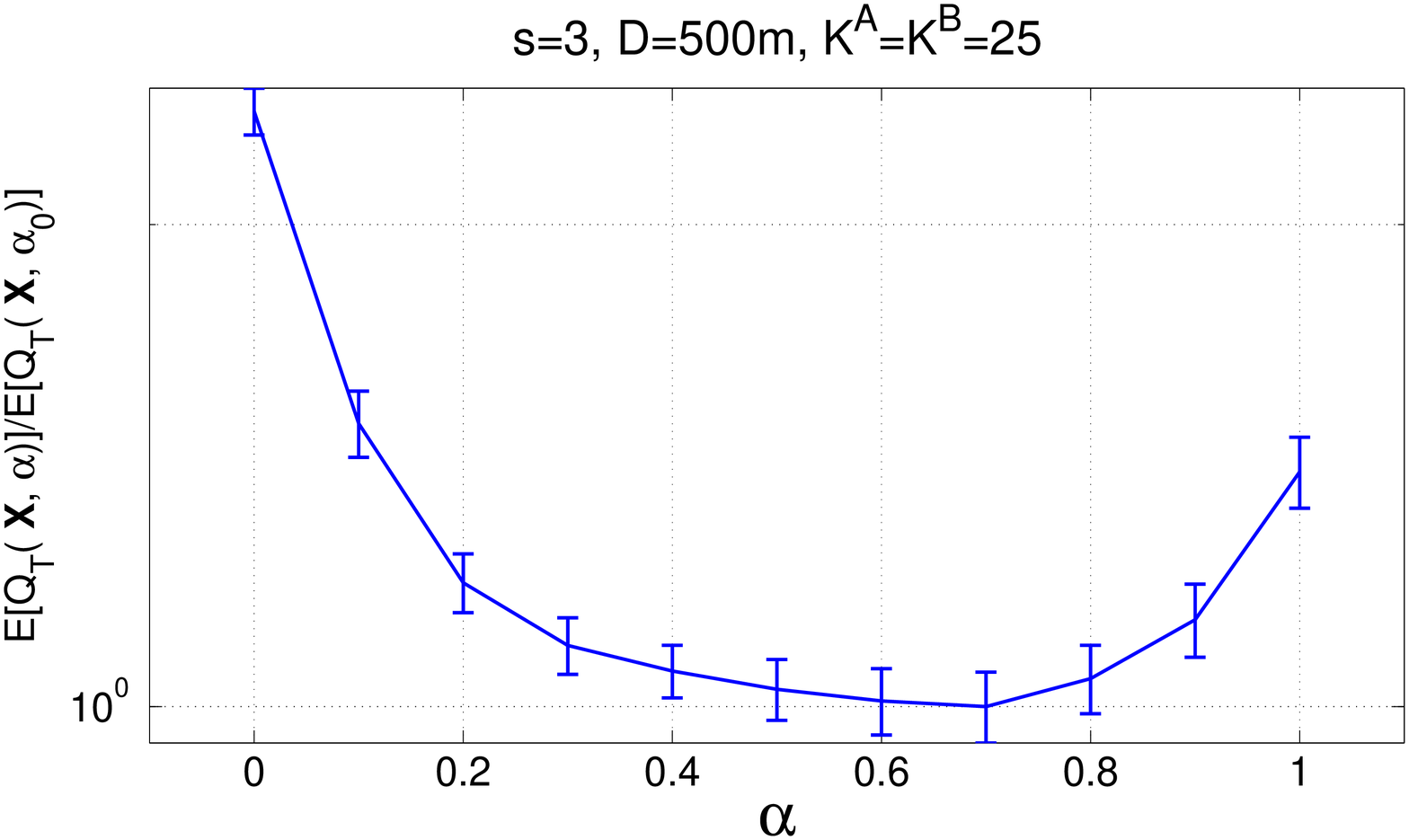}
      \caption{Power vs. $\alpha$ for $s=3$, $D~=~500$ m, $K^A=K^B=25$, $r_t=5$
Mbps}
      \label{fig:alphaQ2}
   \end{minipage}\hfill
   \begin{minipage}[b]{0.49\linewidth}   
      \centering
\includegraphics[scale=0.25]{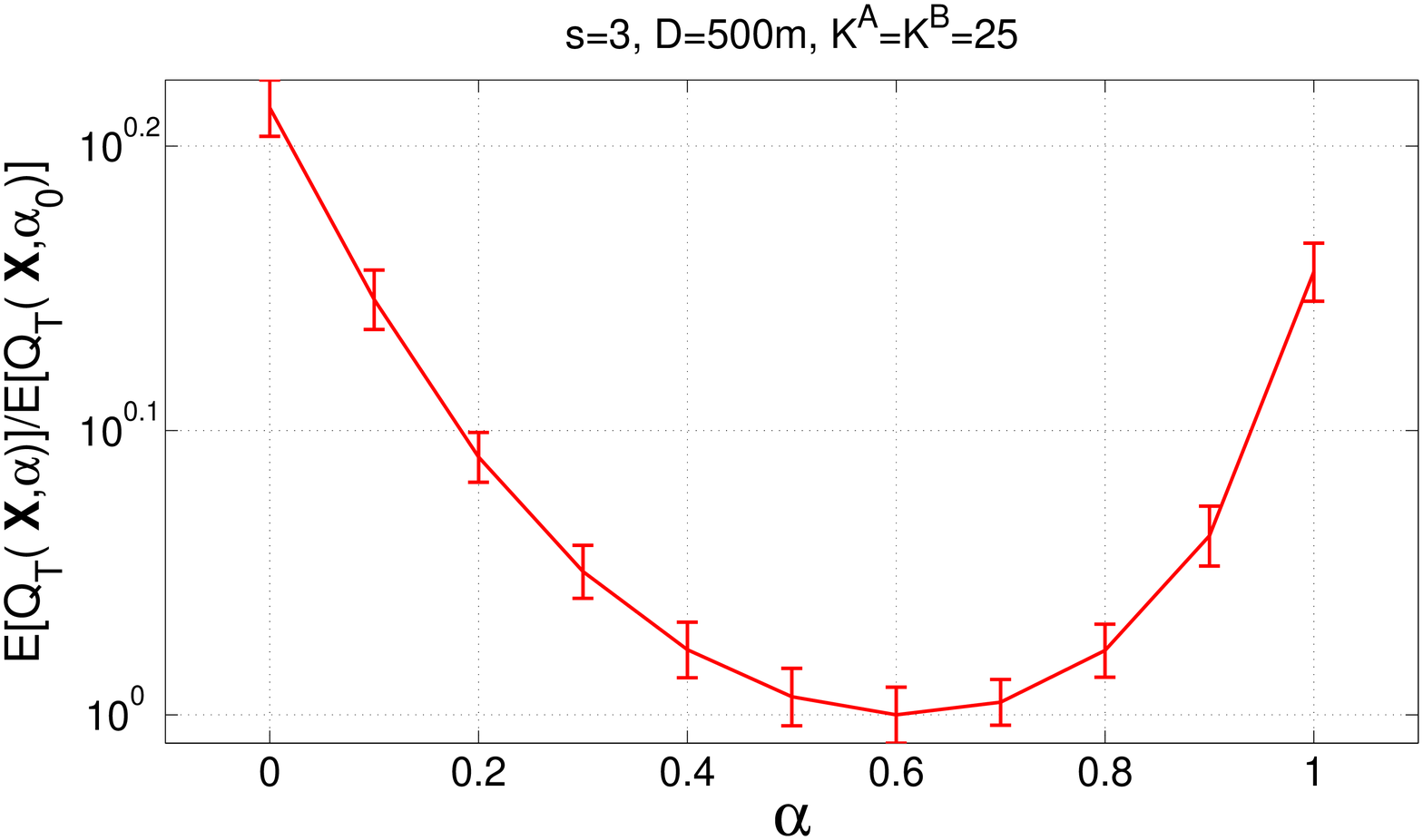}
      \caption{Power vs. $\alpha$ for $s=3$, $D~=~500$ m, $K^A=K^B=25$, $r_t=10$
Mbps}
      \label{fig:alphaQQ2}
   \end{minipage}
\end{figure}

In our simulations, we considered a Free Space Loss model (FSL) characterized by
a path loss exponent $s=2$ as well as the so-called Okumura-Hata (O-H) model for
open areas \cite{okumura} with a path loss exponent $s=3$. 
The carrier frequency is $f_0=2.4 GHz$. At this frequency, path loss in dB 
is given by $\rho_{dB}(x) = 20\log_{10}(x) + 100.04$ in the case where $s=2$,
where $x$ is the distance in kilometers between the BS and the user. In the case
$s=3$, $\rho_{dB}(x) = 30\log_{10}(x) + 97.52$. The signal bandwidth $B$ is
equal to $5$ MHz and the thermal noise power spectral density is equal to $N_{0}
= -170$ dBm/Hz. Each cell has a radius $D = 500$m and contains the same number
of randomly distributed users ($K^A = K^B$). The rate requirement of user $k$ in
bits/sec/Hz is designated by $R_k$. The distance separating each user from the
base station is considered a random variable with a uniform distribution on the
interval~$[0,D]$. The joint resource allocation problem for Cells~$A$ and~$B$
(Problem~\ref{prob:multi}) was solved for a large number of realizations of this
random distribution of users and the values of the resulting transmit power were
averaged. Computing the mean value of the total transmit power with respect to
the random positions of users is intended to get results that do not depend on
the particular position of each user in the cell but rather on global
information about the geographic distribution
of users in the cell. We give now more details on the way simulation were
carried out.\\
Define $\mathbf{x}$ as the vector containing the positions of all the users in
the system \emph{i.e}, $\mathbf{x}=(x_1,x_2,\ldots$, $x_{K^c})_{c=A,B}$. Recall
that $\forall k$, $x_k$ is a random variable with a uniform distribution on
$[0,D]$. For each realization of $\mathbf{x}$, denote by
$Q_T(\mathbf{x},\alpha)$ 
the minimal total transmit power that results from a global solution to the
multicell resource allocation problem (Problem~\ref{prob:multi}) \emph{i.e.,}
$Q_T(\mathbf{x},\alpha)=\sum_{c=A,B}\left(\sum_{k=1}^{L^c}W\ku^c+\sum_{k=L^c}^{
K^c}W\kd^c\right)$
where $(\gamma\ku^c,W\ku^c,\gamma\kd^c,W\kd^c)_{c\in\{A,B\},k=1,\ldots,K^c}$ is
a global solution to Problem~\ref{prob:multi} described by
Theorem~\ref{the:multi}. Define $r_t=\sum_{k=1}^{K^c}R_k B$ as the sum rate of
the users of Cell~$c$ measured in bits/sec. 
\begin{figure}[h]
   \begin{minipage}[b]{0.49\linewidth}
      \centering \includegraphics[scale=0.25]{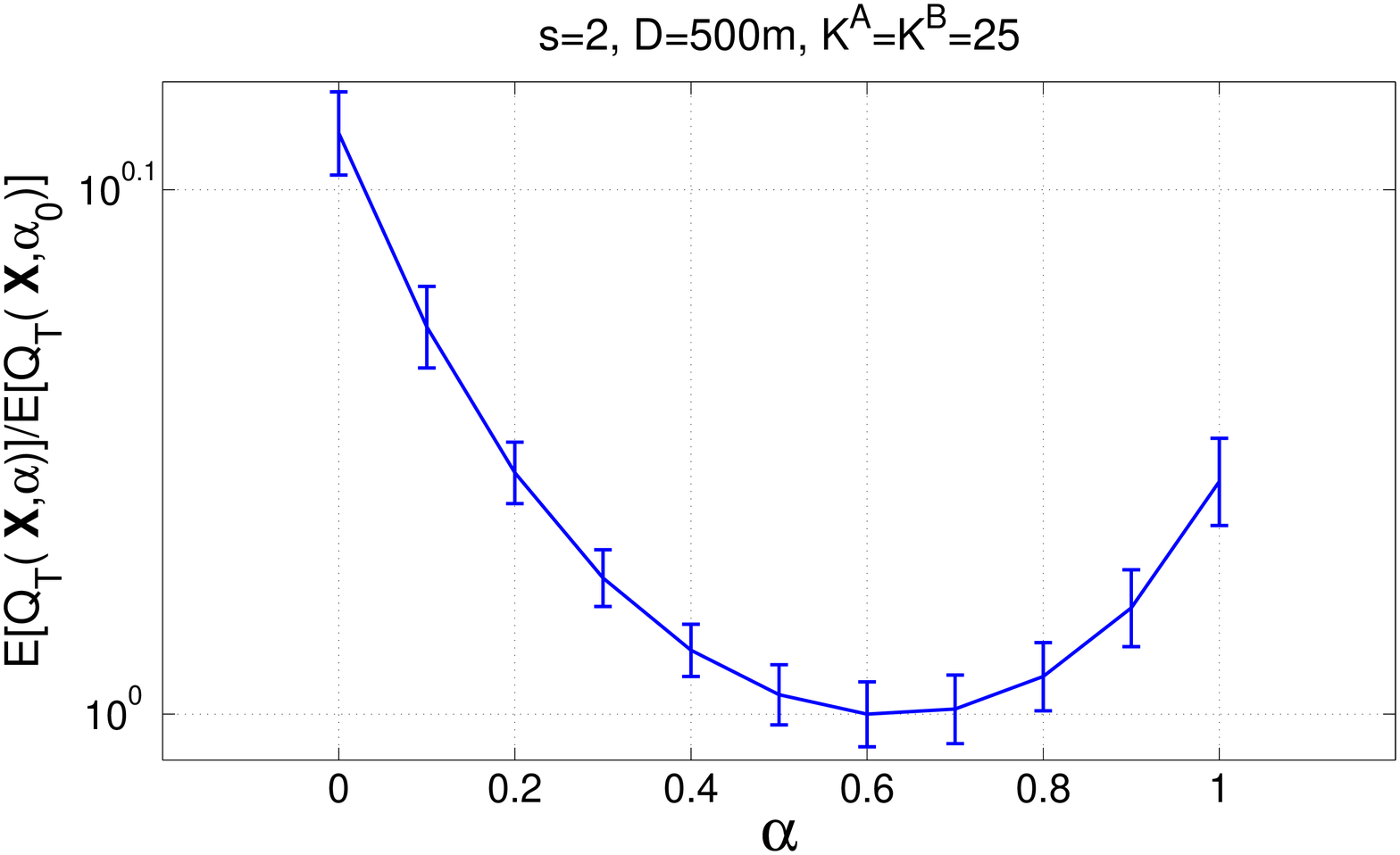}
      \caption{Power vs. $\alpha$ for 
               $s=2$, $D~=~500$ m, $K^A=K^B=25$, $r_t=5$ Mbps}
      \label{fig:alphaQ}
   \end{minipage}\hfill
   \begin{minipage}[b]{0.49\linewidth}   
      \centering \includegraphics[scale=0.25]{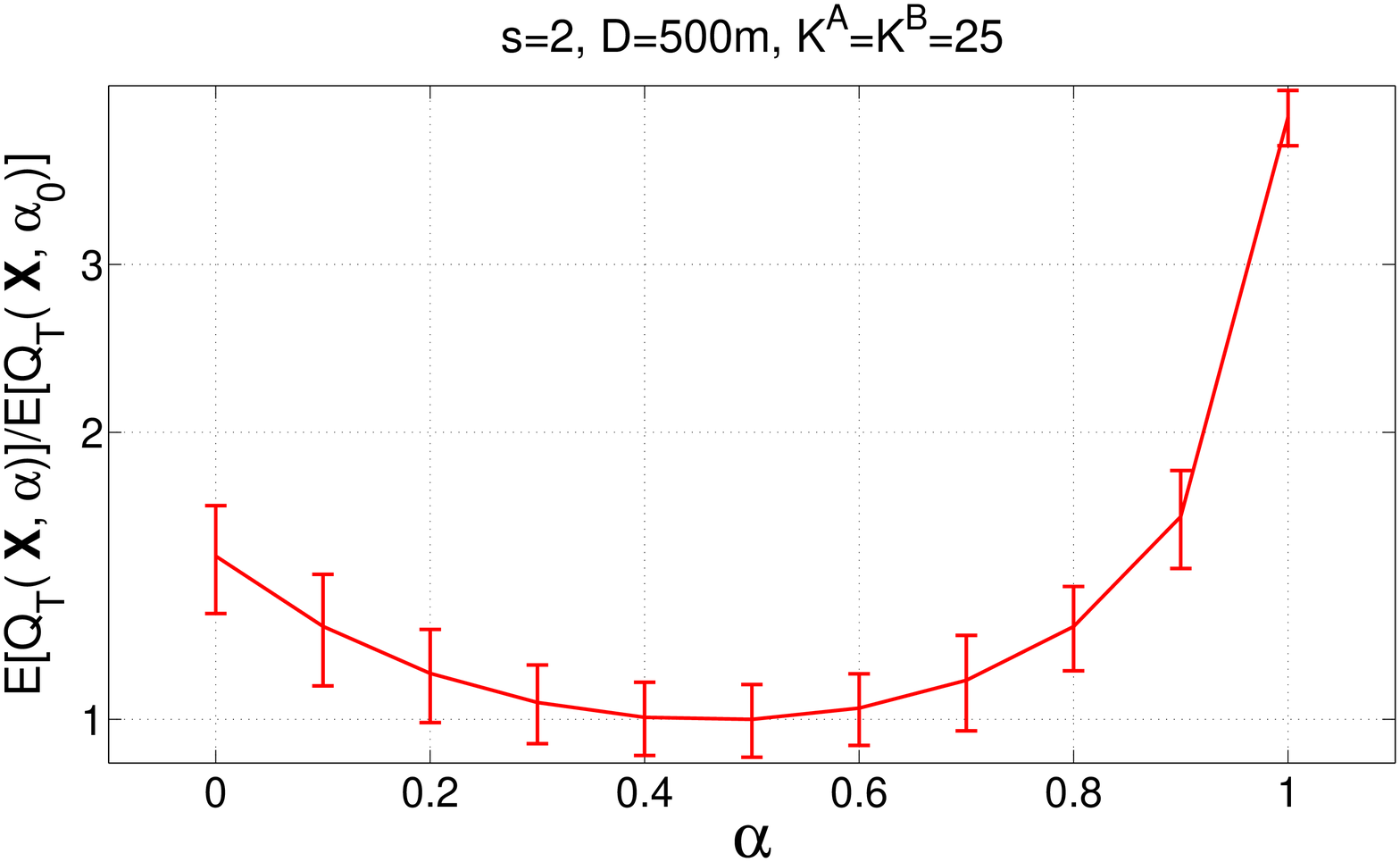}
      \caption{Power vs. $\alpha$ for $s=2$, $D~=~500$ m, $K^A=K^B=25$, $r_t=10$
Mbps}
      \label{fig:alphaQQ}
   \end{minipage}
\end{figure}
We consider first the case where all the users have the same rate requirement
$R_1=R_2=\ldots=R_{K^c}$.
Figures~\ref{fig:alphaQ} and~\ref{fig:alphaQQ}  represent, for a sum rate
requirement of $r_t = 5$ Mbps (Mega bits/sec) and $r_t = 10$ Mbps respectively
and assuming $s=2$, the mean value of $Q_T(\mathbf{x},\alpha)$ normalized by its
minimum value w.r.t $\alpha$ \emph{i.e.,} the ratio
$\mathbb{E}_{\mathbf{x}}[Q_T(\mathbf{x},\alpha)]/\mathbb{E}_{\mathbf{x}}[
Q_T(\mathbf{x},\alpha_0)]$,
where $\alpha_0$ is the value of the reuse factor $\alpha$ that minimizes 
$\mathbb{E}[Q_T(\mathbf{x},\alpha)]$. Figures~\ref{fig:alphaQ2}
and~\ref{fig:alphaQQ2} plot the same quantity for $r_t = 5$ Mbps and $r_t = 10$
Mbps respectively, but with the difference that it assumes $s=3$. The error bars
in the aforementioned four figures represent the variance of
$Q_T(\mathbf{x},\alpha)$ \emph{i.e.,} 
$\mathbb{E}_{\mathbf{x}}[\left(Q_T(\mathbf{x},\alpha)-
 \mathbb{E}_{\mathbf{x}}[Q_T(\mathbf{x},\alpha)]\right)^2]$.

For each value of $\mathbf{x}$ and of the reuse factor $\alpha$, 
$Q_T(\mathbf{x},\alpha)$ was computed using the optimal resource allocation
algorithm of Section~\ref{sec:multicell}. Power gains are considerable compared
to the extreme cases $\alpha = 0$ (the available bandwidth is shared in an
orthogonal way between Cells~$A$ and~$B$) and $\alpha=1$ (all the available
bandwidth is reused in the two cells). Note also that for $r_t=10$ Mbps,
$\alpha_0$ the optimal value of the reuse factor that minimizes
$Q_T(\mathbf{x},\alpha)$ is smaller than the optimal value of the reuse factor
for $r=5$ Mbps. This result is expected, given that higher values of $r_t$ will
lead to higher transmit powers in order to satisfy users' rate requirements, and
consequently to higher levels of interference. More users will need thus to be 
protected from the higher interference. For that purpose, a larger part of the
available bandwidth must be reserved for the protected bands~$\mathcal{P}_A$ and
$\mathcal{P}_B$.
We also remark that in the case where $s=3$, the value of the reuse factor
$\alpha_0$ is larger than its value for $s=2$. This is due to the fact that when
the path loss exponent is larger, the interference produced by the adjacent base
station will undergo more fading than in the case when the path loss exponent is
smaller. As a result, less users need to be protected from interference in the
case $s=3$ compared to the case $s=2$. (see table~\ref{tab:per} which provides,
in the two cases, the percentage of protected users to the total number of users
for $r_t=5$ and $r_t=10$ Mbps, provided that the corresponding value of
$\alpha_0$ is used in each case). 
\begin{table}[h]\label{tab:per}
\begin{center}
\begin{tabular}{l|l|l}
~& $s=2$ & $s=3$\\ 
\hline
$r_t=5$ Mbps& $19.8\%$ & $11.6\%$\\ 
\hline
$r_t=10$ Mbps& $30.0\%$ & $18.7\%$\\
\end{tabular}
\caption{Percentage of the protected users to the total number of users}
\end{center}
\end{table}

\begin{figure}[h]
   \begin{minipage}[b]{0.49\linewidth}
      \centering \includegraphics[scale=0.25]{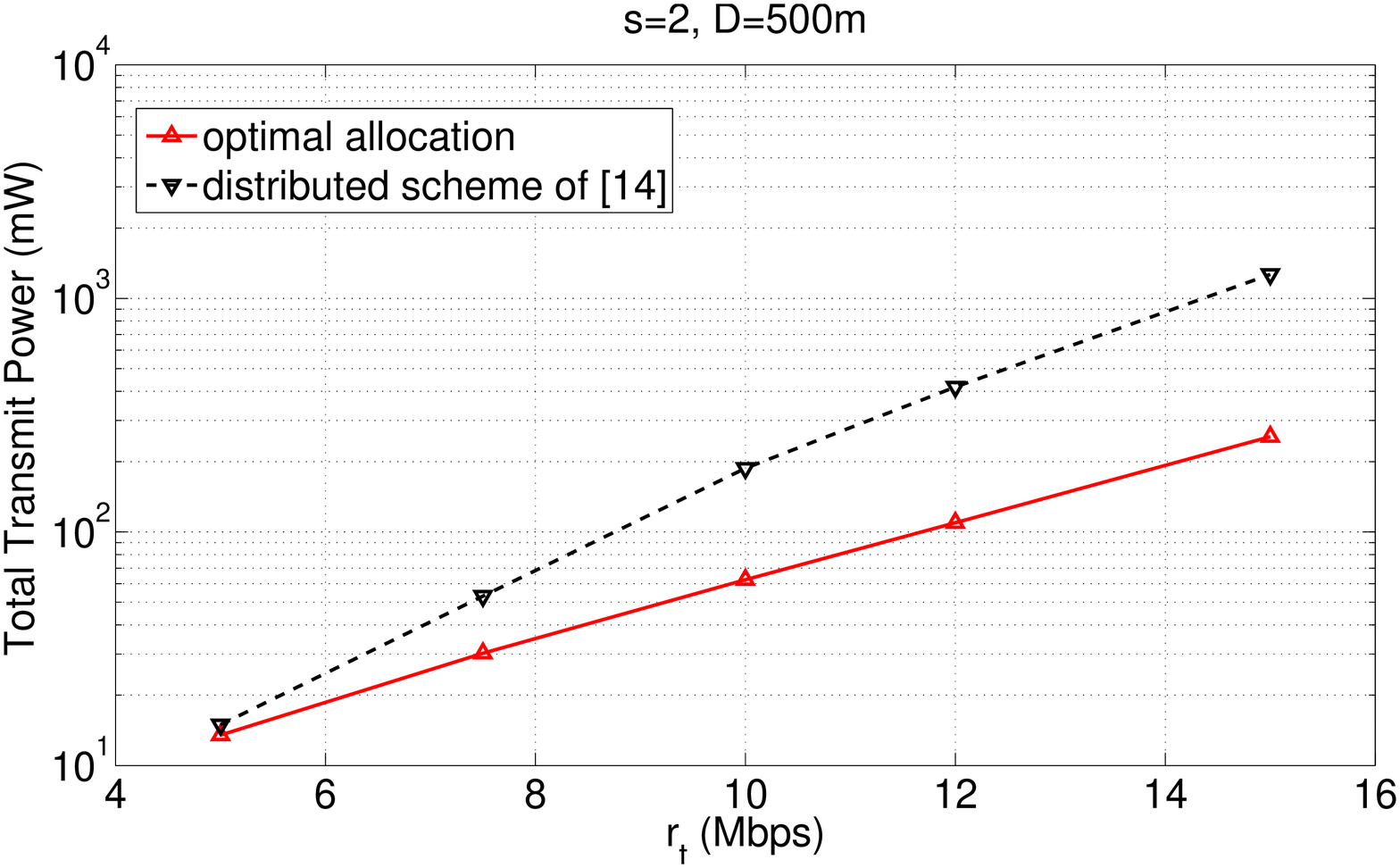}
      \caption{Comparison between the proposed optimal scheme and the
distributed scheme of~\cite{papandriopoulos} for $K^A=K^B=25$.}
      \label{fig:opt_yates}
   \end{minipage}\hfill
   \begin{minipage}[b]{0.49\linewidth}   
      \centering
\includegraphics[scale=0.25]{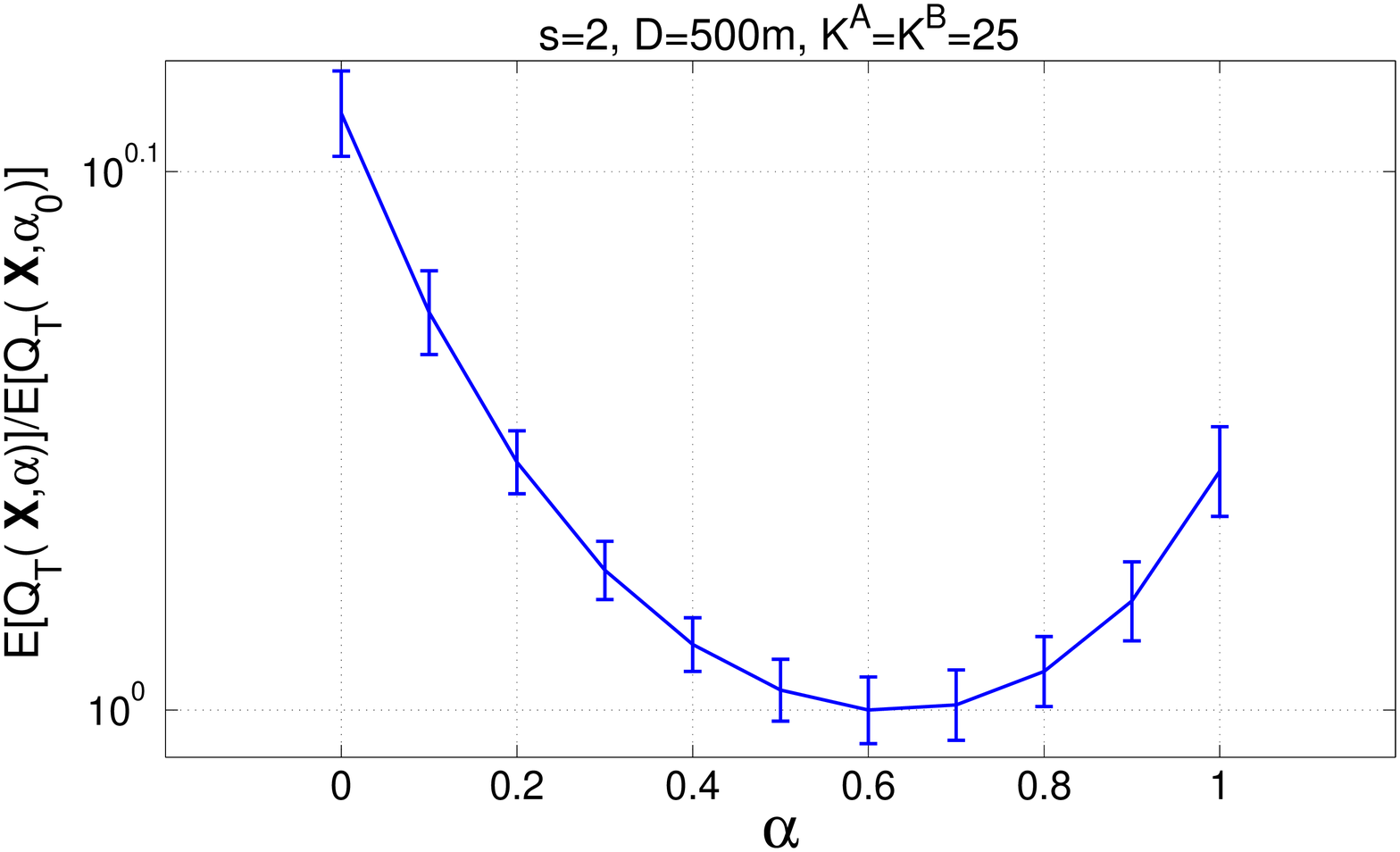}
      \caption{Power vs. $\alpha$ for $s=2$, $D~=~500$ m, $K^A=K^B=25$ assuming
random rate requirements.}
      \label{fig:alphaQ3}
   \end{minipage}
\end{figure}

We now compare the performance of our proposed resource allocation with the
distributed scheme proposed in~\cite{papandriopoulos}. The latter scheme assumes
a reuse factor $\alpha$ equal to one (all the subcarriers can be reused in all
the cells), in contrast to our scheme which uses an optimized value of $\alpha$.
Figure~\ref{fig:opt_yates} plots the average total transmit power
$\mathbb{E}[Q_T^{(K)}({\bf X},\alpha_{0})]$ that results when our proposed
scheme is applied compared to the power that results from applying the scheme
of~\cite{papandriopoulos}. This comparison was carried out assuming
$K^A=K^B=25$, $s=2$ and $r_t=5$ Mbps. The gain obtained when the proposed scheme
is applied is clear from the figure, and it increases with respect to the total
rate $r_t$.
%
We consider now the case when the rate requirement is not the same for all
users. In particular, we assume that the rate requirement of each user is a
random variable that can take on one of two values with the same probability.
For example, consider the case $K^A=K^B=25$ and assume that the rate requirement
of each user can either be equal to $250$ kbps (kilo bits/sec) with probability
$0.5$ or to $150$ kbps with the same probability. This means that the mean rate
per user is equal to $200$ kbps and that the mean total rate per sector is equal
to $r_t=25*200$ kbps = 5 Mbps. 
Figure~\ref{fig:alphaQ3} represents, assuming $s=2$, the mean value of
$Q_T(\mathbf{x},\alpha)$ normalized by its minimum value w.r.t $\alpha$
\emph{i.e.,} the ratio
$\mathbb{E}_{\mathbf{x}}[Q_T(\mathbf{x},\alpha)]/\mathbb{E}_{\mathbf{x}}[
Q_T(\mathbf{x},\alpha_0)]$,
where $\alpha_0$ is the value of the reuse factor $\alpha$ that minimizes
$\mathbb{E}[Q_T(\mathbf{x},\alpha)]$.
The error bars in the above figure represent the variance of
$Q_T(\mathbf{x},\alpha)$ \emph{i.e.,} 
$\mathbb{E}_{\mathbf{x}}[\left(Q_T(\mathbf{x},\alpha)-\mathbb{E}_{\mathbf{x}}[
Q_T(\mathbf{x},\alpha)]\right)^2]$. 
By comparing Figures~\ref{fig:alphaQ} and~\ref{fig:alphaQ3} 
we note that the normalized mean value
$\mathbb{E}_{\mathbf{x}}[Q_T(\mathbf{x},\alpha)]/\mathbb{E}_{\mathbf{x}}[
Q_T(\mathbf{x},\alpha_0)]$ is practically the same in the two figures. 
Only the variance
$\mathbb{E}_{\mathbf{x}}[\left(Q_T(\mathbf{x},\alpha)-
\mathbb{E}_{\mathbf{x}}[Q_T(\mathbf{x},\alpha)]\right)^2]$ is slightly different
(its value is slightly larger in the case of random rate requirements).
\section{Conclusions}

In this paper, the resource allocation problem for a sectorized downlink OFDMA
system has been studied in the context of a partial reuse factor 
$\alpha\in[0,1]$. The general solution to the (nonconvex) optimization problem
has been provided. It has been proved that the solution admits a simple form and
that the initial tedious problem reduces to the identification of a restricted
number of parameters. As a noticeable property, it has been proved that the
optimal resource allocation policy is ``binary'': 
there exists a pivot-distance to the base station such that users who are
farther than this distance should only modulate protected subcarriers, while
closest users should only modulate reused subcarriers. 
\appendices
\section{Proof of Theorem~\ref{the:single}}
\label{app:main_theo}

When the resource allocation parameters of users in Cell $B$ are fixed, it is
straightforward to show that the ergodic capacity $C_k=C_k(\gamma\ku^A,
\gamma\kd^A, W\ku^A, W\kd^A)$ defined by~(\ref{eq:ergodic_capacity})
is a concave function of $\gamma\ku^A$, $\gamma\kd^A$, $W\ku^A$, $W\kd^A$ (and
hence $-C_k(\gamma\ku^A, \gamma\kd^A, W\ku^A, W\kd^A)$ is convex). This is
essentially due to the fact that $g\ku=g\ku(Q_1^B)$ can be treated as a constant
and does not depend on the optimization parameters. Thus, the single cell
resource allocation problem (Problem~\ref{prob:single}) is convex in
$\{\gamma\ku^A, \gamma\kd^A, W\ku^A, W\kd^A\}_{k\in\{1,\ldots,K^A\}}$. In the
following, we derive the KKT conditions in order to obtain the general form of
the solution and to prove the existence of $L$, $\beta_1$, $\beta_2$, $\xi$ as
stated by Theorem~\ref{the:single}. In particular, we prove that any optimal
resource allocation is binary \emph{i.e.}, there exists a certain pivot-integer
$L$ such that $\gamma\kd^A=0$ for $k<L$ and $\gamma\ku^A=0$ for $k>L$.
Furthermore, we prove that there exist three parameters $\beta_1$, $\beta_2$ and
$\xi$ such that equations~(\ref{eq:allocinf}),~(\ref{eq:allocsup})
and~(\ref{eq:allocL}) hold. As explained above, $\beta_1, \beta_2, \xi$ are the
Lagrange multipliers associated with constraints $\bf C2$, $\bf C3$ and $\bf C6$
respectively.\\
{\bf KKT Conditions for Problem~\ref{prob:single}}

\begin{figure}[h]
   \begin{minipage}[b]{0.49\linewidth}
      \centering \includegraphics[scale=0.25]{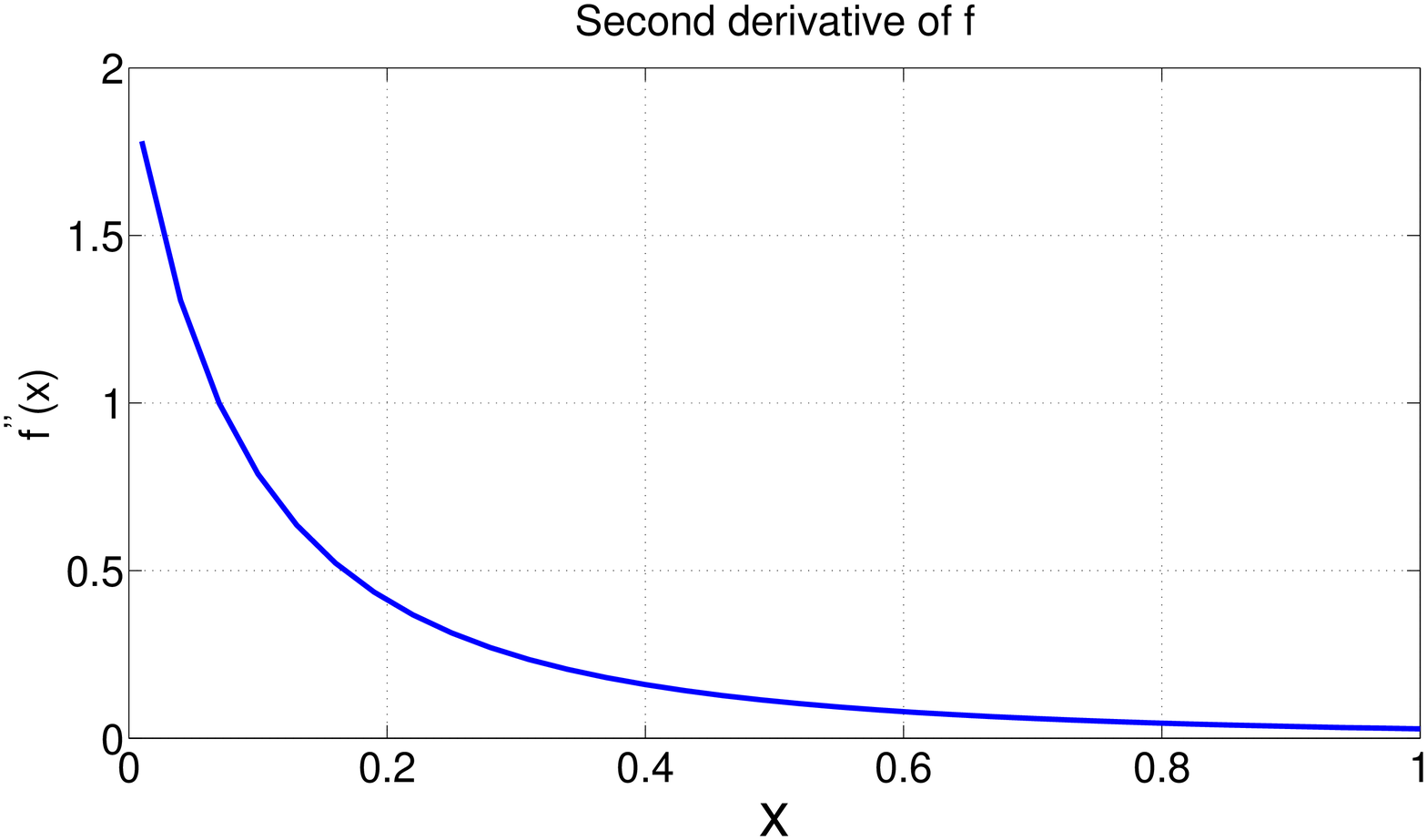}
      \caption{Second derivative of function $f$}
      \label{fig:conj_f}
   \end{minipage}\hfill
   \begin{minipage}[b]{0.49\linewidth}   
      \centering \includegraphics[scale=0.25]{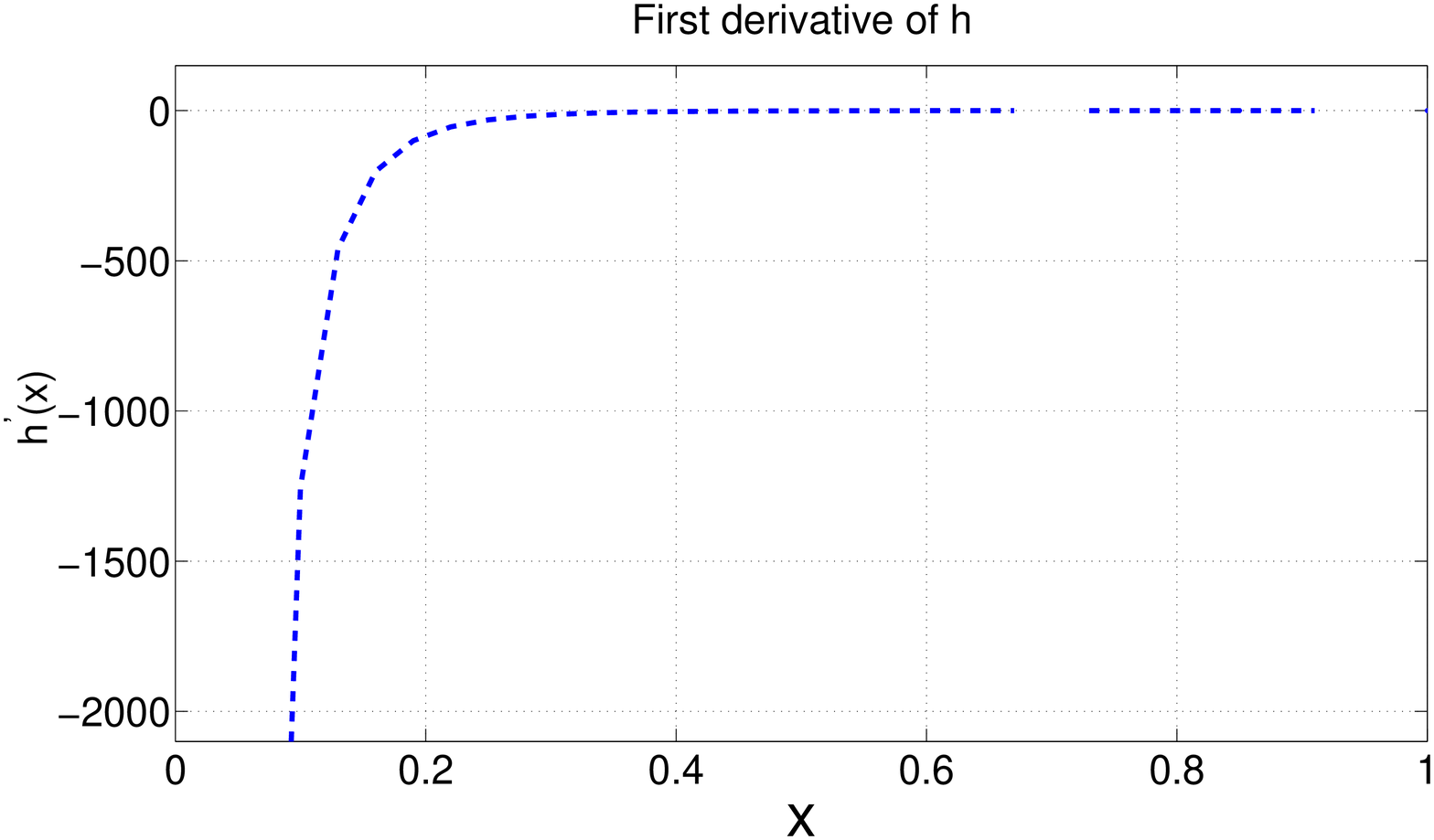}
      \caption{First derivative of function $h$}
      \label{fig:conj_h}
   \end{minipage}
\end{figure}

In order to simplify the notations and since we are only interested in users of
Cell $A$, we simply omit the superscript $A$ in the sequel and define $Q=Q^A$,
$\gamma\ku=\gamma\ku^A$, etc. Denote by $\mathbf{x}_A$ the vector of resource
allocation parameters of users in Cell $A$ \emph{i.e,}
$\mathbf{x}_A=[(\mathbf{W})^T,(\boldsymbol{\gamma})^T]^T$ where
$\mathbf{W}=[W_{1,1},W_{1,2},\ldots,W_{K^A,1},W_{K^A,2}]^{T}$ and
$\boldsymbol{\gamma}=[\gamma_{1,1},\gamma_{1,2},\ldots,\gamma_{K^A,1},\gamma_{
K^A,2}]^{T}$.
The associated Lagrangian is equal to: 
\begin{align}\label{eq:lagrangian}
\mathcal{L}&=  Q-\sum_{k=1}^{K^A}\lambda_k C_k 
+\beta_1\left(\sum_{k}\gamma\ku\right)+\beta_2\left(\sum_{k}\gamma\kd\right)  -
\nonumber \\
&\sum_{k}\nu\ku\gamma\ku-\sum_{k}\nu\kd\gamma\kd  - \sum_{k}\mu\ku
W\ku-\sum_{k}\mu\kd W\kd
+ \xi \sum_{k}W\ku.
\end{align}
where $\lambda_k$, $\beta_1$, $\beta_2$ and $\xi$ are the Lagrange multipliers
associated respectively with constraints $\mathbf{C1}$, $\mathbf{C2}$,
$\mathbf{C3}$ and $\bf C6$ of Problem~\ref{prob:single},
and where $\nu\ku,\nu\kd,\mu\ku,\mu\kd$ are the the Lagrange multipliers
associated with the positivity constraints of $\gamma\ku,\gamma\kd,W\ku,W\kd$
respectively. In the expression of $C_k$, a technical difficulty arises from the
fact that function
$\gamma\ki\EE{\log\left(1+g\ki\frac{W\ki}{\gamma\ki}Z\right)}$ is not
differentiable at 
point $\gamma\ki=0$. One can easily overcome this issue by replacing the
non-negativity constraint $\gamma\ki\geq 0$
by the strict positivity constraint $\gamma\ki\geq \epsilon_0$, for an arbitrary
$\epsilon_0>0$. However, as this point is essentially technical, we simply put
$\epsilon_0=0$ with slight lack of rigor. This assumption will simplify the
presentation without changing the results. The complete proof that does not make
this simplifying assumption can be found in~\cite{rapport}. 
We now apply the Lagrange-Karush-Kuhn-Tucker conditions to characterize the
optimal vector ${\bf x}_A$. Taking the derivative of
\eqref{eq:lagrangian} with respect to $W\ki$ and $\gamma\ki$ ($i=1,2$) leads to
\begin{align}
1-\lambda_k
g\ki\mathbb{E}\left[\frac{Z}{1+g\ki\frac{W\ki}{\gamma\ki}Z}\right]
-\mu\ki+\xi\delta_i&=0 \label{eq:cons_kkt1}\\
-\lambda_{k}\mathbb{E}\left[\log\left(1+g\ki\frac{W\ki}{\gamma\ki}Z\right)-\frac
{g\ki\frac{W\ki}{\gamma\ki}Z}{1+g\ki\frac{W\ki}{\gamma\ki}Z}\right]+\beta_i
-\nu\ki&=0\label{eq:cons_kkt2}
\end{align}
where $\delta_i=1$ if $i=1$ and $\delta_i=0$ if $i=2$. We can easily show that
the constraint $R_k\leq C_k$ must hold with equality, and is always active in
the sense that the Lagrange multiplier $\lambda_k$ associated with this
constraint is strictly positive. Identifying parameter $\lambda_k$
in~(\ref{eq:cons_kkt1}) and~(\ref{eq:cons_kkt2}) yields
$f\left(g\ki\frac{W\ki}{\gamma\ki}\right)=\frac{g\ki(\beta_i
-\nu\ki)}{1-\mu\ki+\xi\delta_i}$,
where $f$ is the function defined by~\eqref{eq:functions_fF}. Replacing the
value of $g\ki\frac{W\ki}{\gamma\ki}$ in~(\ref{eq:cons_kkt1}) by
$f^{-1}\left(\frac{g\ki(\beta_i -\nu\ki)}{1-\mu\ki+\xi\delta_i}\right)$ directly
provides the following equation:
\begin{equation}
\label{eq:result_kkt1}
1-\mu\ki+\xi\delta_i=\lambda_k g\ki F\left(\frac{g\ki(\beta_i
-\nu\ki)}{1-\mu\ki+\xi\delta_i}\right),
\end{equation}
where $F$ is the function defined by~\eqref{eq:functions_fF2}. Define
$\mathcal{A}_i=\{k/\nu\ki=0\}$. In other words, $\mathcal{A}_1$ is the set of
users of Cell $A$ being assigned non zero share of the band $\cal I$, and
$\mathcal{A}_2$ is the set of users of Cell $A$ being assigned non zero share of
$\mathcal{P}_A$. By complementary slackness, we may write on the opposite
$\overline{\mathcal{A}_i}=\{k/\gamma\ki=0\}$ where $\overline{E}$ denotes
the complementary set of any set $E\subset \{1,\dots K^A\}$. After some algebra,
it can be shown that $\nu\ki=0$ implies $\mu\ki=0$. Thus, 
\begin{equation}
  \label{eq:cons_KKT_b}
      \forall k \in \mathcal{A}_i, \;\; \frac{g\ki}{1+\xi\delta_i}
F\left(\frac{g\ki}{1+\xi\delta_i} \beta_i\right)=\lambda_k^{-1} \:.
\end{equation}
On the other hand, if $\nu\ki>0$, KKT conditions lead to
\begin{equation}\label{eq:nu_0_inequality}
      \forall k \in \overline{\mathcal{A}_i}, \;\; \frac{g\ki}{1+\xi\delta_i}
F\left(\frac{g\ki}{1+\xi\delta_i} \beta_i\right)<\lambda_k^{-1}
\end{equation}
To prove that inequality~\eqref{eq:nu_0_inequality} holds, one needs to separate
the two possible cases $W\ki=0$ and $W\ki> 0$. \emph{i)} If $W\ki=0$,
equation~\eqref{eq:cons_kkt2} leads to $\beta_i=\nu\ki$. Thus,
(\ref{eq:result_kkt1}) is equivalent to $1-\mu\ki+\xi\delta_i=\lambda_k g\ki$,
which implies that $\frac{g\ki}{1+\xi\delta_i}\leq \lambda_k^{-1}$ since
$\mu\ki\geq 0$. Noticing that $ F\left(\frac{g\ki}{1+\xi\delta_i} \beta_i\right)
< 1$ and multiplying this inequality by the previous one, we obtain the
desired equation~(\ref{eq:nu_0_inequality}). \emph{ii)} If $W\ki>0$,
complementary slackness condition $\mu\ki W\ki=0$ along with
equation~(\ref{eq:result_kkt1}) lead to $\mu\ki=0=1+\xi\delta_i-\lambda_k g\ki
F\left(\frac{g\ki(\beta_i -\nu\ki)}{1-\mu\ki+\xi\delta_i}\right)$. As function
$F(x)$ is strictly decreasing,
$F\left(\frac{g\ki}{1+\xi\delta_i} \beta_i\right) < F\left(\frac{g\ki(\beta_i
-\nu\ki)}{1-\mu\ki+\xi\delta_i}\right) =
\frac{1+\xi\delta_i}{\lambda_k g\ki}$. We thus obtain
inequality~\eqref{eq:nu_0_inequality} as well.\\
To summarize, every global solution to our optimization problem can thus be
characterized by the following set of conditions:
\begin{enumerate} 
\item For every $k \in \mathcal{A}_i$:
\begin{align}
  \frac{g\ki}{1+\xi\delta_i}
F\left(\frac{g\ki}{1+\xi\delta_i}\beta_i\right)=\lambda_k^{-1},	
&\quad \frac{W\ki}{\gamma\ki}=
g\ki^{-1}f^{-1}\left(\frac{g\ki}{1+\xi\delta_i}\beta_i\right)\label{eq:kkt_f1}
\end{align}
\item For every $k \in \bar{\mathcal{A}_i}$:
  \begin{align}
    \frac{g\ki}{1+\xi\delta_i}
F\left(\frac{g\ki}{1+\xi\delta_i}\beta_i\right)<\lambda_k^{-1},	
    &\quad W\ki=0\label{eq:kkt_f2}
  \end{align}
\item \begin{equation*}
    \forall k\:\: C_k=R_k, \;\; \quad\sum_{k}\gamma\ku=\alpha, \;\;
\quad\sum_{k}\gamma\kd=\frac{1-\alpha}{2},\;\; 
    \xi \left(\sum_k W\ku -{\cal Q}\right)=0\:.
  \end{equation*}
\end{enumerate}
We determine now which users are in $\mathcal{A}_1$ and which are in
$\mathcal{A}_2$. For that sake, the following conjecture will be revealed useful
in the sequel. Define $h(x)=\frac{x(F^{-1}(x))^{'}}{F^{-1}(x)}$.
\begin{conj}
  Function $f(x)$ is strictly convex. Function $h(x)$ is non increasing on the
interval $(0,1)$.
\label{conj:single_conj}
\end{conj}
In order to validate the above conjecture, Figures~\ref{fig:conj_f}
and~\ref{fig:conj_h} represent the second respectively derivative of $f$ which
is obviously positive, and the first derivative of $h$, which is obviously
negative on $(0,1)$. 
%
%
We show now that equations~(\ref{eq:cons_KKT_b}) and~(\ref{eq:nu_0_inequality})
are sufficient to prove that the following lemma holds.
\begin{lemma}\label{claim:single_binary} Any global solution to
Problem~\ref{prob:single} is ``binary'' \emph{i.e.}, there exists a user $L$ in
Cell $A$ such that $\gamma\kd=0$ for closest users $k<L$, and $\gamma\ku=0$ for
farthest users $k>L$. 
\end{lemma}
\begin{proof}
Now define $L=\min \mathcal {A}_2$ as the closest user to the base station among
all users modulating in the protected band ${\cal P}_A$. By definition of $L$,
we have $\gamma_{1,2} = \dots = \gamma_{L-1,2}=0$ which is equivalent to the
first part of the desired result. Now we prove the second part \emph{i.e.},
$\gamma_{L+1,1} = \dots = \gamma_{K^A,1}=0$. To simplify notations, we define
for each user $k$, $\tilde{g}\ku=\frac{g\ku}{1+\xi}$.
By definition, $L\in {\cal A}_2$. By immediate application of the above KKT
conditions, $g\Ld F(g\Ld\beta_2) = \lambda_k^{-1} \geq  \tilde{g}\Lu
F\left(\tilde{g}\Lu \beta_1\right)$. 
As $F$ is decreasing, we obtain $\beta_2 < \frac{1}{g\Ld}F^{-1}\left(
\frac{\tilde{g}\Lu}{g\Ld} F\left(\tilde{g}\Lu \beta_1\right)\right)$. Now
consider a second user $k\geq L+1$ and assume by contradiction that $k\in{\cal
A}_1$.
Using the same arguments, it is straightforward to show that $\beta_2 >
\frac{1}{g\kd}F^{-1}\left( \frac{\tilde{g}\ku}{g\kd} F\left(\tilde{g}\ku
\beta_1\right)\right)$. Putting all pieces together,
$\frac{1}{g\kd}F^{-1}\left( \frac{\tilde{g}\ku}{g\kd} F\left(\tilde{g}\ku
\beta_1\right)\right)
<\frac{1}{g\Ld}F^{-1}\left( \frac{\tilde{g}\Lu}{g\Ld} F\left(\tilde{g}\Lu
\beta_1\right)\right)$.
We now prove that the above inequality cannot hold when $k>L$. To that end, we
introduce the following notations. Define $x=\tilde{g}\Lu \beta_1$,
$r=\frac{\rho_k}{\rho_L}$, $t=\frac{\sigma_L^2}{\sigma_k^2}$ and
$s=\frac{\sigma^2}{\sigma_L^2(1+\xi)}$. Using these notations, the above
inequality reduces to
\begin{equation}
  \frac{1}{r}F^{-1}\left( st F\left(r t x\right)\right)<F^{-1}\left( s
F\left(x\right)\right)\:.
\label{eq:contra}
\end{equation}
Note that in the above inequality, all variables $r$, $s$, $t$ are strictly less
than one. We now prove with the help of Conjecture~\ref{conj:single_conj} that
the above inequality leads to a contradiction. In fact,
Conjecture~\ref{conj:single_conj} states that function $f(x)$ is strictly
convex. As $f(x)$ is also strictly increasing, its inverse $f^{-1}$ is strictly
concave strictly increasing. Therefore, for every $t<1$ and for every $y>0$,
$f^{-1}(ty) > tf^{-1}(y)$. Using the definition of function $F(x)$, it is
straightforward to show that the latter inequality leads to  
\begin{equation}\label{eq:ineq1}
\forall (r,s,t)\in (0,1)^3,\;\;\;
\frac{1}{r}F^{-1}(stF(trx))>\frac{1}{r}F^{-1}(sF(rx))
\end{equation}
for each real $x$. As function $h(x)=\frac{x(F^{-1}(x))^{'}}{F^{-1}(x)}$ is non
increasing on $(0,1)$,
it can be shown after some algebra~\cite{rapport} that function $r \to
\frac{1}{r}F^{-1}(sF(rx))$ 
is decreasing on $(0,1)$. As a consequence,
\begin{equation}\label{eq:ineq2}
\forall (r,s)\in (0,1)^2,\;\;\; \frac{1}{r}F^{-1}(sF(rx)) \geq F^{-1}(s
F\left(x\right))  \:.
\end{equation}
Clearly, \eqref{eq:ineq1} and \eqref{eq:ineq2} contradict
inequality~\eqref{eq:contra}. This proves the desired lemma.
\end{proof}
Lemma~\ref{claim:single_binary} establishes the ``binary'' property of any
global solution to Problem~\ref{prob:single}. One still needs to prove that
equations~\eqref{eq:allocinf}, \eqref{eq:allocsup} and~\eqref{eq:allocL} hold.
Fortunately, these equations result directly from combining the above claim with
equations~\eqref{eq:kkt_f1} and~\eqref{eq:kkt_f2}.

\section{Proof of Theorem~\ref{the:multi}}\label{sec:multi_proof}

\noindent {\sl Notations.}  
In the sequel, $\mathbf{x}_{AB}$ represents a vector of multicell allocation
parameters such that ${\bf x}_{AB}=[{{\bf x}_A}^T,{{\bf x}_B}^T]^T$ where
$\mathbf{x}_A=[(\mathbf{W}^A)^T,(\boldsymbol{\gamma}^A)^T]^T$ and
$\mathbf{x}_B=[(\mathbf{W}^B)^T,(\boldsymbol{\gamma}^B)^T]^T$ and where  for
each $c=A,B$, $\mathbf{W}^c=[W_{1,1}^c,W_{1,2}^c,\ldots,
W_{K^c,1}^c,W_{K^c,2}^c]^{T}$ and
$\boldsymbol{\gamma}=[\gamma_{1,1}^c,\gamma_{1,2}^c,\ldots,\gamma_{K^c,1}^c,
\gamma_{K^c,2}^c]^{T}$. We respectively denote by
$Q_1(\mathbf{x}_c)=\sum_k
W\ku^c$ and $Q_2(\mathbf{x}_c)=\sum_kW\kd^c$ the powers transmitted by Base
Station~$c$ in the interference band $\cal I$ and in the protected band ${\cal
P}_c$. When resource allocation ${\bf x}_{AB}$ is used, the total power
transmitted by the network is equal to 
$Q({\bf x}_{AB})=\sum_cQ_1(\mathbf{x}_c)+Q_2(\mathbf{x}_c)$.\\
Recall that Problem~\ref{prob:multi} is nonconvex. It cannot be solved using
classical convex optimization methods.
Denote by ${\bf x}_{AB}^*=[{{\bf x}_A^*}^T,{{\bf x}_B^*}^T]^T$ any global
solution to
Problem~\ref{prob:multi}.
\smallskip

\noindent {\bf Characterizing ${\bf x}_{AB}^*$ via single cell results}.

From ${\bf x}_{AB}^*$ we construct a new vector ${\bf x}_{AB}$ which is as well
a global solution and which admits a ``binary'' form: for each Cell $c$,
$\gamma\ku^c=0$
if $k>L^c$ and $\gamma\kd^c=0$ if $k<L^c$, for a certain pivot-integer $L^c$.
For each Cell $c$, vector ${\bf x}_A$ is defined as a global solution to the
\emph{single cell} allocation Problem~\ref{prob:single} when 
\begin{itemize}
\item[a)] the admissible nuisance constraint $\cal Q$ is set to ${\cal
Q}=Q_1({\bf x}_A^*)$,
\item[b)] the gain-to-interference-plus-noise-ratio in band $\cal I$ is set to
$g\ku=g\ku\left( Q_1({\bf x}_{B}^*) \right)$.
\end{itemize}
Vector ${\bf x}_B$ is defined similarly, by simply exchanging $A$ and $B$ in the
above definition. Denote by ${\bf x}_{AB} =[{{\bf x}_A}^T,{{\bf x}_B}^T]^T$ the
resource allocation obtained by the above procedure. The following Lemma holds.
\begin{lemma}
  Resource allocation parameters ${\bf x}_{AB}$ and ${\bf x}_{AB}^*$ coincide:
${\bf x}_{AB}={\bf x}_{AB}^*$.
\end{lemma}
\begin{proof}
It is straightforward to show that ${\bf x}_{AB}$ is a feasible point for the
joint multicell Problem~\ref{prob:multi} in the sense that constraints $\bf
C1$-$\bf C5$ of Problem~\ref{prob:multi} are met. This is the consequence of the
low nuisance constraint $Q_1({\bf x}_c) \leq Q_1({\bf x}_c^*)$ which ensures
that the interference which is \emph{produced} by each base station when using
the new allocation ${\bf x}_{AB}$ is no bigger than the interference produced
when the initial allocation ${\bf x}_{AB}^*$ is used. Second, it is
straightforward to show that ${\bf x}_{AB}$ is a global solution to the
multicell Problem~\ref{prob:multi}.
Indeed, the power $Q_1(\mathbf{x}_c)+Q_2(\mathbf{x}_c)$ spent by Base
Station~$c$ is necessarily less than the initial power
$Q_1(\mathbf{x}_c^*)+Q_2(\mathbf{x}_c^*)$ \emph{by definition} of the
minimization Problem~\ref{prob:single}. Thus $Q({\bf x}_{AB})\leq Q({\bf
x}_{AB}^*)$. Of course, as ${\bf x}_{AB}^*$ has been chosen itself as a global
minimum of $Q$, the latter inequality should hold with equality: $Q({\bf
x}_{AB})= Q({\bf x}_{AB}^*)$.
Therefore, ${\bf x}_{AB}^*$ and ${\bf x}_{AB}$ are both global solutions to the
multicell
Problem~\ref{prob:multi}. 
As an immediate consequence, inequality $Q_1(\mathbf{x}_c)+Q_2(\mathbf{x}_c)\leq
Q_1(\mathbf{x}_c^*)+Q_2(\mathbf{x}_c^*)$ holds with equality in both Cells $c$:
\begin{equation}
  Q_1(\mathbf{x}_c)+Q_2(\mathbf{x}_c)=
Q_1(\mathbf{x}_c^*)+Q_2(\mathbf{x}_c^*)\:.
\label{eq:xetoilemin}
\end{equation}
Clearly, ${\bf x}_A^*$ is a feasible point for Problem~\ref{prob:single} when
setting constant ${\cal Q} = Q_1({\bf x}_A^*)$ and $g\ku = g\ku\left(Q_1({\bf
x}_B^*)\right)$. Indeed constraint $\bf C6$ is equivalent to $Q_1({\bf x}_A^*)
\leq {\cal Q}$ and is trivially met (with equality) by definition of $\cal Q$.
Since the objective function $Q_1(\mathbf{x}_A^*)+Q_2(\mathbf{x}_A^*)$ coincides
with the global minimum as indicated by~(\ref{eq:xetoilemin}), ${\bf x}_A^*$ is
a global minimum for the single cell Problem~\ref{prob:single}. By
Theorem~\ref{the:single}, this single cell problem admits a unique global
minimum ${\bf x}_A$. Therefore, ${\bf x}_A^*={\bf x}_A$. By similar arguments,
${\bf x}_B^*={\bf x}_B$.
\end{proof}
Using the above Lemma along with Theorem~\ref{the:single}, we conclude that any
global solution ${\bf x}_{AB}^*$ to the joint multicell Problem~\ref{prob:multi}
satisfies equations~\eqref{eq:multiinf}, \eqref{eq:multisup} and
\eqref{eq:multiL}, where parameters $L^c,\beta_1^c,\beta_2^c,\xi^c$ for
${c=A,B}$ in the latter equations can be defined as in
Appendix~\ref{sec:determination_L_beta} using values $g\ku=g\ku\left(Q_1({\bf
x}_{\bar c}^*)\right)$ and ${\cal Q}=Q_1({\bf x}_{c}^*)$. The proof of
Theorem~\ref{the:multi} is thus complete.


\end{document}